\documentclass[acmsmall]{acmart}

\AtBeginDocument{%
  }

\usepackage{ifthen}
\usepackage{xcolor}
\usepackage[most]{tcolorbox}

\newboolean{showcomments}
\setboolean{showcomments}{true} 

\ifthenelse{\boolean{showcomments}}{
	\newcommand{\nbc}[3]{
		{\colorbox{#3}{\bfseries\sffamily\scriptsize\textcolor{white}{#1}}}
		{\textcolor{#3}{\sf\small$\langle$\textit{#2}$\rangle$}}}
	
}{
	\newcommand{\nbc}[3]{}

}

\renewcommand\footnotetextcopyrightpermission[1]{}

\setcopyright{acmlicensed}
\copyrightyear{2026}
\acmYear{2026}
\acmDOI{XXXXXXX.XXXXXXX}

\acmJournal{CSUR}

\begin{document}

\title{Bidirectional Empowerment of Metamorphic Testing and Large Language Models: A Systematic Survey}

\author{Zheng Zheng}
\email{zhengz@buaa.edu.cn}
\orcid{0000-0001-7922-9067}
\author{Zenghui Zhou}
\email{zhouzenghui@buaa.edu.cn}
\author{Yinwang Xu}
  \email{xuyinwang@buaa.edu.cn}
\affiliation{%
  \institution{Beihang University}
  \city{Beijing}
  \country{China}
}

\author{Daixu Ren}
\email{rendaixu@tiangong.edu.cn}
\affiliation{%
  \institution{Tiangong University}
  \city{Tianjin}
  \country{China}}
\authornote{Corresponding author.}

\author{Tsong Yueh Chen}
 \email{tychen@swin.edu.au}
\affiliation{%
 \institution{Swinburne University of Technology}
 \city{Melbourne}
 \country{Australia}}

\renewcommand{\shortauthors}{Zheng et al.}

\begin{abstract}
  Large language models (LLMs) have introduced substantial challenges to software quality assurance due to their generative, probabilistic, and open-ended nature, which intensifies the oracle problem and limits the applicability of traditional testing methods. Metamorphic testing (MT), which checks necessary relations among multiple related executions rather than relying on exact expected outputs, has emerged as a promising approach for testing LLMs and other oracle-deficient systems. At the same time, the strong semantic understanding, reasoning, and code generation capabilities of LLMs create new opportunities to automate the traditionally labor-intensive phases of MT. This survey systematically reviews 93 primary studies and characterizes this reciprocal relationship as the bidirectional empowerment of MT and LLMs. We propose a taxonomy spanning two complementary directions: MT for LLMs, which uses MT to verify, validate, assess, and understand LLMs and LLM-based systems across issues such as hallucination, fairness, robustness, code reliability, retrieval-augmented generation, dialogue, and autonomous agents; and LLMs for MT, which leverages LLMs to support metamorphic relation discovery, input transformation and synthesis, executable test implementation, and agentic closed-loop testing. By synthesizing these developments, this survey provides a structured foundation for understanding the evolving synergy between MT and LLMs and highlights future directions for building more rigorous, scalable, and trustworthy AI quality assurance methodologies.
\end{abstract}

\ccsdesc[500]{Software and its engineering~Software testing and debugging}
\ccsdesc[500]{Software and its engineering~Software verification and validation}
\ccsdesc[300]{Computing methodologies~Natural language processing}
\ccsdesc[300]{Computing methodologies~Artificial intelligence}

\keywords{Metamorphic Testing, Large Language Models, Software Testing, Oracle Problem, Test Automation}


\maketitle

\section{Introduction}

Large language models (LLMs) have rapidly transformed artificial intelligence and software engineering. Built upon large-scale pre-training and transformer architectures~\cite{vaswani2017attention,brown2020language}, LLMs have demonstrated strong capabilities in natural language understanding and generation, complex reasoning~\cite{wei2022chain}, code generation~\cite{hou2024large}, and tool-augmented or agentic workflows~\cite{park2023generative,yao2023react}. These capabilities have accelerated the integration of LLMs into increasingly consequential applications, including autonomous driving~\cite{fu2024drive,baimbetova2026llm}, automated program repair (APR)~\cite{fan2023automated}, medical question answering~\cite{singhal2025toward}, retrieval-augmented generation (RAG), dialogue systems, and autonomous agents. As LLMs move from experimental models to components of real-world systems, their reliability, robustness, fairness, and safety have become central concerns for both research and practice.

Assuring LLM-based systems is, however, fundamentally difficult. Unlike many conventional software systems whose expected outputs can often be specified, computed, or checked against deterministic assertions, LLMs produce probabilistic, open-ended, and semantically rich outputs. For many tasks, such as question answering, code synthesis, dialogue, and agent planning, there may be multiple acceptable outputs rather than a single ground truth. This makes the classical oracle problem~\cite{barr2015oracle} particularly severe in the LLM setting. Moreover, LLMs exhibit failure modes that are difficult to capture with static benchmarks or exact-match assertions, including hallucination~\cite{ji2023survey}, demographic and social bias~\cite{sheng2021societal}, adversarial vulnerability~\cite{zou2023universal}, and non-deterministic behavior. Practitioner-oriented discussions have further emphasized that manually labeled ground-truth evaluation does not scale to the volume and diversity of inputs required for LLM-powered software~\cite{terragni2026untestable}. These characteristics create an urgent need for testing and evaluation methods that can reason about behavioral relations rather than relying exclusively on exact expected outputs.

Metamorphic testing (MT)~\cite{chen1998metamorphic,Chan1998aomt,segura2016survey,chen2018metamorphic} provides a natural response to this challenge. MT was originally proposed as a technique for generating follow-up test cases and has since become a prominent oracle-alleviating testing paradigm for systems whose exact outputs are difficult or impossible to determine. Instead of checking whether a single output is correct, MT checks whether a necessary relation, known as a metamorphic relation (MR), holds among the outputs of multiple related inputs. For example, although it may be difficult to determine the uniquely correct response to a complex prompt, one may still expect the model to preserve sentiment under synonym replacement, maintain factual consistency under paraphrasing, or produce functionally equivalent code under semantics-preserving prompt transformations. Such relational properties make MT especially suitable for LLMs, where correctness is often contextual, semantic, and comparative rather than absolute.

Recent studies have increasingly applied MT to LLMs and LLM-based systems. Existing work has used MT to detect hallucinations in open-domain and RAG systems~\cite{li2024drowzee,wu2025detectinga,sok2025metarag}, uncover fairness violations under demographic perturbations~\cite{reddy2025metamorphic,giramata2025efficient,romero2025meta}, evaluate robustness and consistency under semantics-preserving transformations~\cite{hyun2024metal,cho2025llmorph,decurto2025metamorphic}, and assess the reliability of code generation models~\cite{chan2025effectiveness,wang2024validating,honarvar2025turbulence}. These studies show that MT can provide a scalable and flexible assurance mechanism for LLMs when traditional test oracles are unavailable or prohibitively expensive.

At the same time, the relationship between MT and LLMs is not one-directional. While MT offers an oracle-alleviating framework for testing LLMs, LLMs also provide new opportunities to automate MT itself. A long-standing obstacle to the broader adoption of MT is the manual effort required to identify effective MRs, construct valid input transformations, and implement executable test workflows~\cite{li2025metamorphic}. The semantic understanding, reasoning, and code-generation capabilities of LLMs can help alleviate this bottleneck. Recent studies have used LLMs to discover MRs from requirements, documentation, and source code~\cite{shin2024towards,srinivas2023potential,xu2026mrcoupler}, synthesize input transformation functions from existing test suites~\cite{xu2024mr,xu2025automated}, and support agentic testing workflows in which test generation, execution, checking, and refinement are performed iteratively~\cite{liang2025automt,gogani2025llm,canizares2026llmmtworkflows}. Consequently, LLMs are not only systems under test, but also assistants or agents for scaling the MT lifecycle.

\begin{figure}[h]
  \centering
  \includegraphics[width=\linewidth]{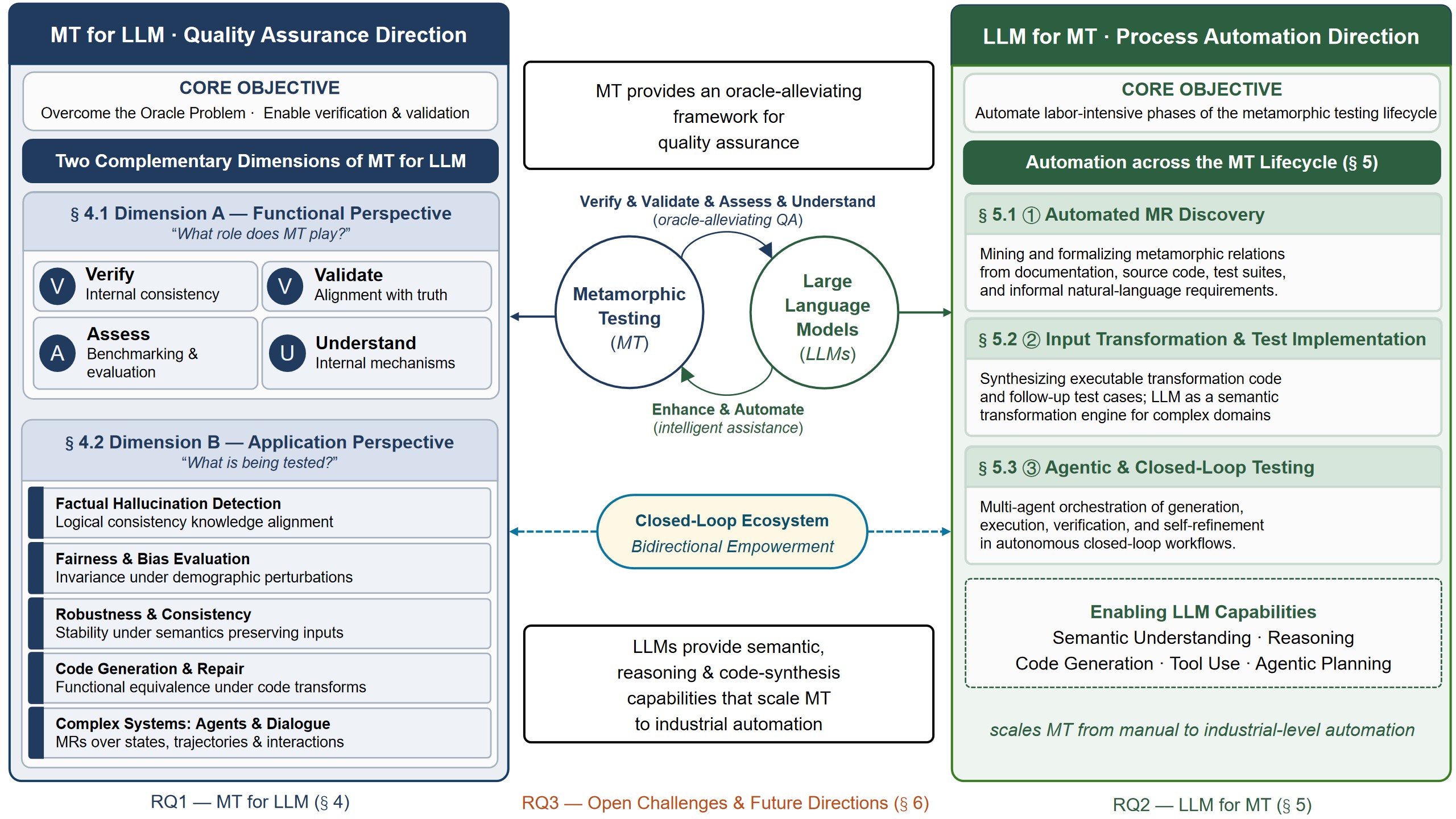}
  \caption{The bidirectional empowerment of MT and LLMs}
  \label{fig:empowerment}
\end{figure}
\begin{figure}
    \centering
    \includegraphics[width=0.8\linewidth]{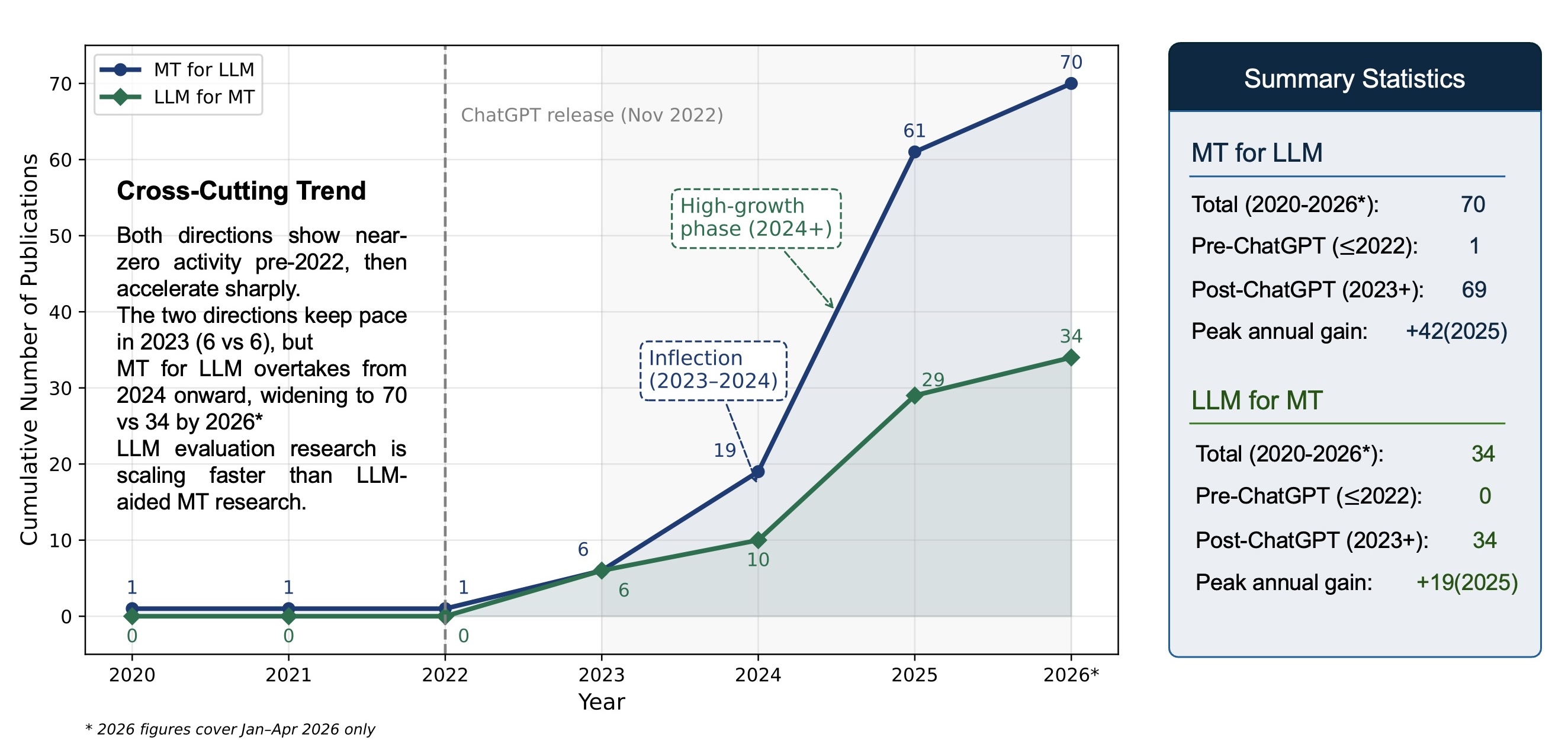}
    \caption{Cumulative publications from 2020 to 2026 for the two directions.}
    \label{fig:Publications}
    
    \vspace{2pt}
    \footnotesize{\textit{Note:} The counts are direction-wise and non-exclusive; studies addressing both directions are counted in both curves.}
\end{figure}

This survey characterizes the emerging interaction between these two areas as the \emph{Bidirectional Empowerment of MT and LLMs}. As shown in Figure~\ref{fig:empowerment}, this perspective consists of two complementary directions. In the first direction, \emph{MT for LLMs}, MT supports the assurance of LLMs and LLM-based systems by enabling relation-based verification, validation, assessment, and behavioral understanding under oracle scarcity. This direction covers concrete quality concerns such as hallucination, fairness, robustness, code-related reliability, RAG, dialogue, and autonomous agents. In the second direction, \emph{LLMs for MT}, LLMs support the automation of the MT lifecycle by assisting in MR discovery, input transformation and synthesis, executable test implementation, and agentic closed-loop testing. Together, these two directions form a closed-loop ecosystem: MT strengthens the trustworthiness of LLMs, while LLMs improve the scalability, usability, and automation of MT.

The need for a systematic synthesis is further evidenced by the rapid growth of the literature. Figure~\ref{fig:Publications} shows the cumulative publication trend from 2020 to 2026. Before the widespread adoption of ChatGPT and related instruction-following LLMs, studies at the intersection of MT and LLMs were relatively sparse. After the release of ChatGPT in late 2022, the field reached a clear inflection point: research on both MT for LLMs and LLMs for MT began to accelerate. The increase became particularly pronounced after 2024, when the number of publications grew rapidly in both directions. This trend indicates that the field has moved beyond isolated proof-of-concept studies toward a more active and coherent research area. It also highlights the timeliness of a dedicated survey that consolidates existing evidence, clarifies terminology, and provides a structured research map.

In this paper, we conduct a systematic review on the interaction between MT and LLMs. The main contributions of this survey are summarized as follows:

\begin{itemize}
    \item We provide, to the best of our knowledge, the first dedicated systematic survey on the bidirectional relationship between MT and LLMs, establishing a unified perspective on this emerging research area.

    \item We propose a structured taxonomy that organizes existing studies into two complementary directions: \emph{MT for LLMs} and \emph{LLMs for MT}.

    \item We provide an evidence-based synthesis of 93 primary studies, revealing how current research
is distributed across functional objectives, application scenarios, and automation phases, and
highlighting the methodological patterns that distinguish the two directions.

    \item We identify major research trends, methodological limitations, and open challenges, and outline future directions toward rigorous, scalable, and trustworthy AI quality assurance.
\end{itemize}

The remainder of this paper is organized as follows. Section~2 introduces the necessary background on MT, LLMs, and their bidirectional synergy. Section~3 presents the systematic review methodology, including the search strategy, study selection criteria, data extraction process, and proposed taxonomy. Section~4 reviews the literature on \emph{MT for LLMs}, organized by functional objectives and target application scenarios. Section~5 surveys the literature on \emph{LLMs for MT}. Section~6 discusses open challenges and future research directions. Finally, Section~7 concludes the paper.

\section{Background and Core Concepts}

\subsection{ Metamorphic Testing}
Traditional software testing often relies on a test oracle, i.e., a mechanism that determines whether the output of a program for a given input is correct. However, for many real-world systems—such as machine learning models, optimization algorithms, and especially LLMs—the expected output may be unknown, ambiguous, or prohibitively expensive to obtain. This challenge is widely referred to as the oracle problem~\cite{barr2015oracle}.
Metamorphic Testing, originally proposed by Chen~\cite{chen1998metamorphic,Chan1998aomt}, is an oracle-alleviating testing technique that verifies program correctness by checking relations among multiple but related executions, rather than validating individual outputs against a ground truth. Instead of asking “Is this output correct?”, MT asks, “Do the outputs produced for related inputs satisfy a necessary property of the intended functionality?”

The core component of MT is the metamorphic relation (MR). Let $f: X \rightarrow Y$ denote the function implemented by the system under test (SUT), where $X$ is the input space and $Y$ is the output space. An MR describes a necessary property of $f$ over a sequence of $n \ge 2$ inputs $x_1, x_2, \dots, x_n \in X$ and their corresponding outputs $f(x_1), f(x_2), \dots, f(x_n)$. Formally, an MR can be defined as a relation:
\[
R \subseteq X^n \times Y^n
\]
where $X^n$ and $Y^n$ denote the Cartesian products of the input and output spaces, respectively. We use 
\[
R(x_1, x_2, \dots, x_n, f(x_1), f(x_2), \dots, f(x_n))
\]
to indicate that the inputs and outputs jointly satisfy the metamorphic relation $R$.

Crucially, an MR specifies a necessary condition that must hold if the SUT is correct; violation of the MR indicates a fault, even when the exact correctness of individual outputs cannot be determined.

In practice, the $n$ inputs involved in an MR are divided into two categories:
\begin{itemize}
    \item Source inputs: $x_1, x_2, \dots, x_k$, where $1 \leq k < n$;
    \item Follow-up inputs: $x_{k+1}, x_{k+2}, \dots, x_n$.
\end{itemize}

The follow-up inputs are constructed from the source inputs (and, if necessary, their outputs) according to the transformation specified by the MR.

Based on this, a Metamorphic Group (MG) of inputs is the ordered sequence of inputs $\langle x_1, x_2, \dots, x_n \rangle$ consisting of both source inputs and their corresponding follow-up inputs. Thus, an MG represents the basic execution unit of MT: the SUT is executed on all inputs in the group, and the MR is checked over the resulting outputs.

Consider an LLM-based sentiment analysis system $f$ that is intended to map an input sentence to a sentiment label in $\{\text{positive, neutral, negative}\}$. A reasonable metamorphic relation for $f$ can be defined as:
Replacing a sentiment-preserving word with its synonym should not change the predicted sentiment label. Accordingly, the source input and follow-up input can be designed as:
\begin{itemize}
    \item Source input $x_1$: The movie is good.
    \item Follow-up input $x_2$: The movie is excellent.
\end{itemize}

Formally, this MR can be expressed as:
\[
f(x_1) = f(x_2)
\]

Here $\langle x_1, x_2 \rangle$ forms an MG. If the SUT produces different sentiment labels for $x_1$ and $x_2$, the MR is violated, indicating a potential reliability issue --- even though we never explicitly verified whether either label is ``correct''.

A typical MT process consists of the following phases:
\begin{enumerate}
    \item \textbf{Source Test Generation}: Generating one or more source inputs $x_1, x_2, \dots, x_k$, using some test case selection methods, such as random selection from existing datasets, or applying domain knowledge.
    \item \textbf{Input Transformation}: Constructing follow-up inputs $x_{k+1}, x_{k+2}, \dots, x_n$ according to the MR (e.g., synonym replacement, paraphrasing, input scaling). It should be noted that the construction of follow-up inputs may need the outputs of the source inputs.
    \item \textbf{Execution}: Running the SUT on all inputs in the metamorphic group.
    \item \textbf{Relation Checking}: Checking whether the outputs satisfy the MR. Any violation is treated as a test failure and  may require further analysis.
\end{enumerate}

A key advantage of MT is that it allows for the generation of massive amounts of test cases without manual checking, as the verification relies on the relation among executions rather than a ground truth oracle.

\subsection{Large Language Models}

LLMs are neural language models trained on massive text corpora to model token sequences and generate fluent natural language. Modern LLMs are largely built upon the Transformer architecture, which replaces recurrent sequence modeling with self-attention mechanisms and enables efficient modeling of long-range dependencies~\cite{vaswani2017attention}. Scaling model size, data, and computation has been shown to improve language modeling performance following empirical scaling laws~\cite{kaplan2020scaling}, and large autoregressive models such as GPT-3 demonstrate strong zero-shot, one-shot, and few-shot generalization through in-context learning~\cite{brown2020language}.

Beyond pre-training, instruction tuning and reinforcement learning from human feedback have become important techniques for aligning LLM behavior with user intent. InstructGPT, for example, fine-tunes GPT-3 using human demonstrations and preference-based reinforcement learning, showing that models trained with human feedback can better follow user instructions and are preferred by human evaluators over larger non-aligned models~\cite{ouyang2022training}. Prompting techniques further extend the capabilities of LLMs without changing model parameters. Chain-of-thought prompting encourages models to generate intermediate reasoning steps, improving performance on arithmetic, symbolic, and commonsense reasoning tasks~\cite{wei2022chain}. A related zero-shot variant shows that simply adding prompts such as ``Let's think step by step'' can elicit multi-step reasoning in sufficiently large models~\cite{kojima2022large}.

Recent LLM systems are also augmented with external knowledge and tools. RAG combines parametric generation with non-parametric retrieval, enabling models to access external documents for knowledge-intensive tasks~\cite{lewis2020retrieval}. Tool-use methods extend this idea by allowing language models to call external APIs or interact with environments. Toolformer trains models to decide when and how to invoke tools such as calculators, search engines, and translation systems~\cite{schick2023toolformer}, while ReAct interleaves reasoning traces with actions so that language models can reason, plan, and interact with external information sources~\cite{yao2023react}. These developments make LLMs not only text generators but also components of broader interactive, retrieval-augmented, and tool-using AI systems.


Despite these remarkable capabilities, LLMs also introduce fundamental challenges to software quality assurance. Unlike traditional software systems whose behaviors are typically governed by explicit logic and deterministic execution, LLMs are probabilistic, data-driven, and highly context-sensitive. Their outputs are often open-ended rather than uniquely determined, and their quality frequently depends on semantic appropriateness, factual alignment, and user intent rather than exact matching against a predefined expected result. As a result, conventional testing practices that rely on precise test oracles, exact and deterministic outputs, and specification-complete assertions are often inadequate.

Several characteristics of LLMs are especially relevant in this regard.
\begin{itemize}
    \item \textbf{Non-determinism and stochastic generation}: LLMs generate outputs by modeling token probabilities, and therefore even identical prompts may lead to different responses under different decoding strategies or runtime conditions. Moreover, non-determinism may persist even when using deterministic decoding settings (e.g., very low temperature or greedy decoding that always selects the highest-probability token), due to implementation- and hardware-level factors. This weakens the assumptions of reproducibility and exact regression comparability that many traditional testing methods depend upon.
    
    \item \textbf{Open-endedness and oracle scarcity}: For many LLM tasks, such as summarization, question answering, dialogue, code generation, multimodal understanding, or reasoning, there may be multiple acceptable outputs of varying style, structure, modality and level of detail. Consequently, correctness is often not absolute but relative, contextual, and partially subjective. This makes it difficult to construct complete and reliable test oracles in the conventional sense.
    
    \item \textbf{Hallucination and factual unreliability}: LLMs may produce fluent and convincing outputs that are factually unsupported, logically inconsistent, or contradicted by the input context. Such hallucinations may manifest as fabricated facts, erroneous reasoning chains, or unjustified citations, and are particularly problematic in high-stakes domains such as healthcare, law, finance, and scientific assistance.

    \item \textbf{Sensitivity to superficial perturbations}: Although LLMs often appear semantically powerful, their behaviors may still change unexpectedly under paraphrasing, formatting variation, prompt restructuring, minor lexical edits, or modality-specific perturbations such as visually irrelevant changes. This raises concerns regarding robustness, consistency, and the reliability of deployed LLM-based systems under realistic usage conditions.

    \item \textbf{Implicit and under-specified requirements}: In many practical applications, the expected behavior of an LLM is described only at a high level, such as being helpful, harmless, unbiased, or faithful to retrieved evidence. Such requirements are difficult to formalize into exact assertions, which further complicates verification and validation.
\end{itemize}

These characteristics collectively make LLMs representative oracle-deficient systems. Accordingly, testing LLMs requires methods that do not depend exclusively on exact expected outputs, but instead can evaluate whether model behavior remains reasonable, stable, and aligned under related conditions. This is precisely where Metamorphic Testing becomes particularly valuable.

\subsection{The Interaction: A Bidirectional Synergy}

The relationship between MT and LLMs is not unidirectional; rather, it is increasingly characterized by a reciprocal and mutually reinforcing synergy. On the one hand, the intrinsic properties of LLMs—such as stochasticity, open-ended generation, implicit specifications, and severe oracle scarcity—make them particularly suitable targets for MT. On the other hand, the semantic understanding, reasoning, and code generation capabilities of LLMs provide new opportunities to automate and scale some traditionally labor-intensive tasks of MT. This survey conceptualizes this reciprocal relationship as the bidirectional empowerment of MT and LLMs.

\vspace{0.5em}
 \textbf{(1) MT for LLMs: Using MT to assure LLMs.} The first direction concerns the application of MT to the verification, validation, assessment, and understanding of LLMs and LLM-based systems. Because many LLM tasks lack a single correct answer, traditional testing is often ineffective or prohibitively expensive. MT addresses this limitation by checking whether necessary relations hold across multiple related executions. Instead of asking whether one response is exactly correct, MT asks whether the model behaves consistently and appropriately when the input is transformed in a way that should preserve—or predictably change—some property of the output.

This makes MT especially useful for examining key quality concerns of modern LLM systems,
such as hallucination, fairness, robustness, and code reliability, across application settings
including RAG, multi-turn dialogue, and autonomous agents.

\vspace{0.5em}
 \textbf{(2) LLMs for MT: Using LLMs to automate Metamorphic Testing.} The second direction concerns the use of LLMs to facilitate the MT process itself. Although MT is conceptually powerful, its broader adoption has long been constrained by the difficulty of identifying effective MRs, constructing valid input transformations, and implementing executable test workflows. These activities often require significant manual effort and substantial domain expertise.

Recent advances in LLMs provide a promising way to alleviate this automation bottleneck. Given natural language documentation, source code, existing test suites, or informal requirements, LLMs can assist in inferring candidate MRs, generating source and follow-up test cases, synthesizing transformation code, and even orchestrating iterative testing loops as autonomous agents. In other words, LLMs can act not only as systems under test, but also as intelligent assistants for the tester. This capability expands the practical applicability of MT in several ways. LLMs can help reduce the manual cost of MR discovery, improve the diversity and naturalness of transformed inputs, generate executable test artifacts, and support agentic workflows in which test generation, execution, checking, and refinement are performed in a closed loop.

 \textbf{(3) A closed-loop ecosystem.} Taken together, these two directions form a closed-loop ecosystem. MT provides LLMs with a rigorous and scalable testing paradigm suited to oracle-deficient, probabilistic, and open-ended systems. Meanwhile, LLMs provide MT with powerful automation capabilities that reduce human effort and broaden its applicability to increasingly complex domains. Thus, MT strengthens the trustworthiness of LLMs, while LLMs strengthen the scalability and usability of MT.

This bidirectional perspective, summarized at a high level in Figure~\ref{fig:empowerment}, serves as the conceptual foundation of the survey. It also motivates the more detailed taxonomy proposed in Section 3, where the two directions are further decomposed into functional objectives, application scenarios, and automation phases.

\section{Methodology and Taxonomy}

To provide a comprehensive and systematic  overview of the emerging synergy between MT and LLMs, we conducted a systematic literature review following the guidelines proposed by Kitchenham et al.~\cite{kitchenham2004procedures}. In structuring our reporting framework and taxonomy, we drew inspiration from foundational surveys in the field of Metamorphic Testing~\cite{chen2018metamorphic, segura2016survey,li2025metamorphic} as well as recent systematic reviews on LLMs~\cite{hou2024large, augusto2025large}.

Our survey is guided by three core research questions (RQs) that reflect the bidirectional nature of the relationship between MT and LLMs. These questions structure the analysis presented in the subsequent sections:

\begin{itemize}

    \item \textbf{RQ1 (MT for LLMs): }How is MT applied to LLMs to achieve the functional objectives of verification, validation, assessment, and understanding, and how are these applied to evaluate key quality attributes such as factual hallucination, fairness, robustness, code reliability, and complex agentic systems? This question systematically explores the dual pathways of applying MT to LLMs. The first part investigates its four core functional roles (corresponding to Section 4.1), while the second part examines how these functions are instantiated to evaluate concrete quality attributes and application scenarios (corresponding to Section 4.2).
    
    \item \textbf{RQ2 (LLMs for MT):} How can the generative and reasoning capabilities of LLMs be leveraged to enhance the applicability and cost-effectiveness of MT, in particular, on the automation aspects of MT? This question explores the role of LLMs as intelligent agents in automating labor-intensive phases, including MR discovery, input transformation, executable test implementation, and closed-loop test refinement. (Answered in Section 5)
    
    \item \textbf{RQ3 (Challenges):} What are the open challenges and future opportunities in this closed-loop ecosystem? This question identifies critical bottlenecks, such as oracle validity and data leakage, and outlines a roadmap for future research. (Answered in Section 6)
\end{itemize}

\subsection{Search Strategy and Data Sources}

Given the rapid evolution of LLMs, we adopted a hybrid search strategy to capture both established peer-reviewed literature and cutting-edge preprints.

\begin{itemize}
    \item \textbf{Data Sources}: We searched five primary scholarly databases and repositories: ACM Digital Library, IEEE Xplore, SpringerLink, ScienceDirect, and arXiv. The inclusion of arXiv is crucial to cover recent developments in LLMs that have not yet completed the traditional publication cycle. Additionally, to ensure the completeness of our survey and mitigate potential indexing gaps in specific repositories, we conducted supplementary searches on Google Scholar. For Google Scholar, we inspected the first 250 results sorted by relevance for each query, because later results were largely irrelevant.
    
    \item \textbf{Search String}: We constructed a Boolean search string to identify the intersection of the two domains. The query used was: \\
     {("Metamorphic Testing" OR "Metamorphic Relation" OR "Metamorphic Group" OR "Metamorphic Oracle" OR "Metamorphic Transformation" OR "Metamorphic Evaluation" OR "Metamorphic Specification" OR "Metamorphic Prompt") AND
     ("Large Language Model" OR "LLM" OR  "GPT"  OR "DeepSeek"  OR "Claude" OR "Gemini" OR "Code Generation" OR "Code Model" OR "Retrieval-Augmented Generation" OR "RAG"  OR "LLM Agent" OR "AI Agent")}
    
    \item \textbf{Search Scope}: To ensure high relevance and precision, we restricted the search to the Title, Abstract, and Keywords fields (metadata). We deliberately excluded full-text searches to filter out studies that merely mention these terms in passing---such as in the Background or Related Work sections---without making them a central focus of the research. To mitigate the potential loss of recall caused by metadata-only search, we complemented the database search with backward and forward snowballing from the selected primary studies and relevant surveys.
    
    \item \textbf{Time Frame}: Our literature search was conducted up to April 30, 2026. While no specific start date was imposed during the initial retrieval to ensure inclusivity, the selected primary studies predominantly span from January 2019 to April 2026. This timeframe corresponds with the emergence of modern LLMs in 2019, which triggered a significant increase in research exploring the synergy between MT and generative AI.
\end{itemize}

\subsection{Study Selection Criteria}

To ensure the relevance and high quality of our systematic survey, we performed literature selection following dual inclusion-exclusion criteria.
Studies were first retained only when complying with all inclusion requirements. Subsequently, qualified papers were further excluded if matching any exclusion criterion.
The full details of screening criteria are presented below:

\textbf{Inclusion Criteria.}
A study was included in this survey only if it satisfied all of the following criteria:
\begin{enumerate}
\item It explicitly adopts MT, defines or uses MRs, or employs metamorphic transformations as a core testing or evaluation mechanism.
\item It targets LLMs or LLM-based systems as the SUT, or leverages LLMs to support or automate metamorphic testing.
\item It is written in English.
\end{enumerate}

\textbf{Exclusion Criteria.}
A study that passed the inclusion screening was excluded from this survey if it met any of the following criteria:
\begin{enumerate}
\item It is a duplicate of another included work; only the most recent or complete version is retained.
\item It uses the acronym “MR” to refer to unrelated concepts (e.g., mixed reality, medical records) rather than metamorphic relations.
\item It focuses on general deep learning or machine learning testing without specific involvement of LLMs, foundation models, or LLM-based applications.
\end{enumerate}

\begin{figure}
    \centering
    \includegraphics[width=1\linewidth]{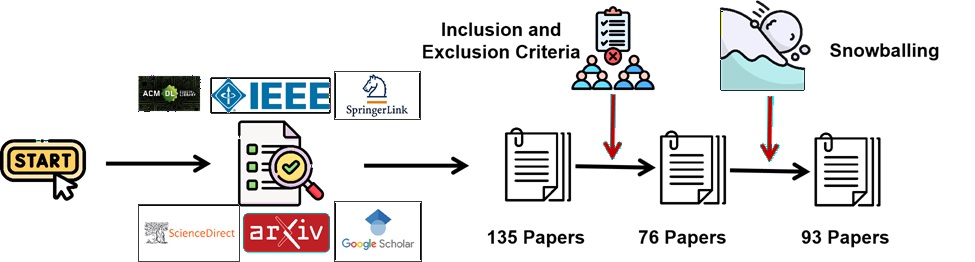}
    \caption{Overview of the paper collection and selection process}
    \label{fig:collectionprocess}
\end{figure}

\textbf{Selection Results}: Figure \ref{fig:collectionprocess} summarizes the overall paper collection and selection process. The
initial database and supplementary searches yielded potentially relevant records. After removing clearly irrelevant records based on metadata screening, 135 papers were
retained for detailed eligibility assessment. We then applied the inclusion and exclusion criteria
described above to the titles, abstracts, keywords, and, where necessary, full texts of these papers.
This stage resulted in 76 papers that directly satisfied our selection criteria. To improve coverage
and reduce the risk of missing relevant studies, we further conducted backward and forward
snowballing on the retained papers and relevant surveys. Snowballing added 17 additional studies,
leading to a final set of 93 primary studies included in this survey.

\subsection{Quality Assessment and Data Extraction}

For each selected study, we extracted bibliographic information, publication year, publication venue, publication type, target system, role of MT, role of LLMs, MR type, input transformation strategy, output checking mechanism, evaluated models, datasets, metrics, and reported limitations. We then coded each study according to the  taxonomy proposed in the next subsection. Because a study may contribute to more than one category, multi-label coding was allowed. To reduce subjective bias, each paper was first coded by one author and then cross-checked by at least one other author. Disagreements were resolved through discussion until consensus was reached.

\begin{figure}
    \centering
    \includegraphics[width=\linewidth]{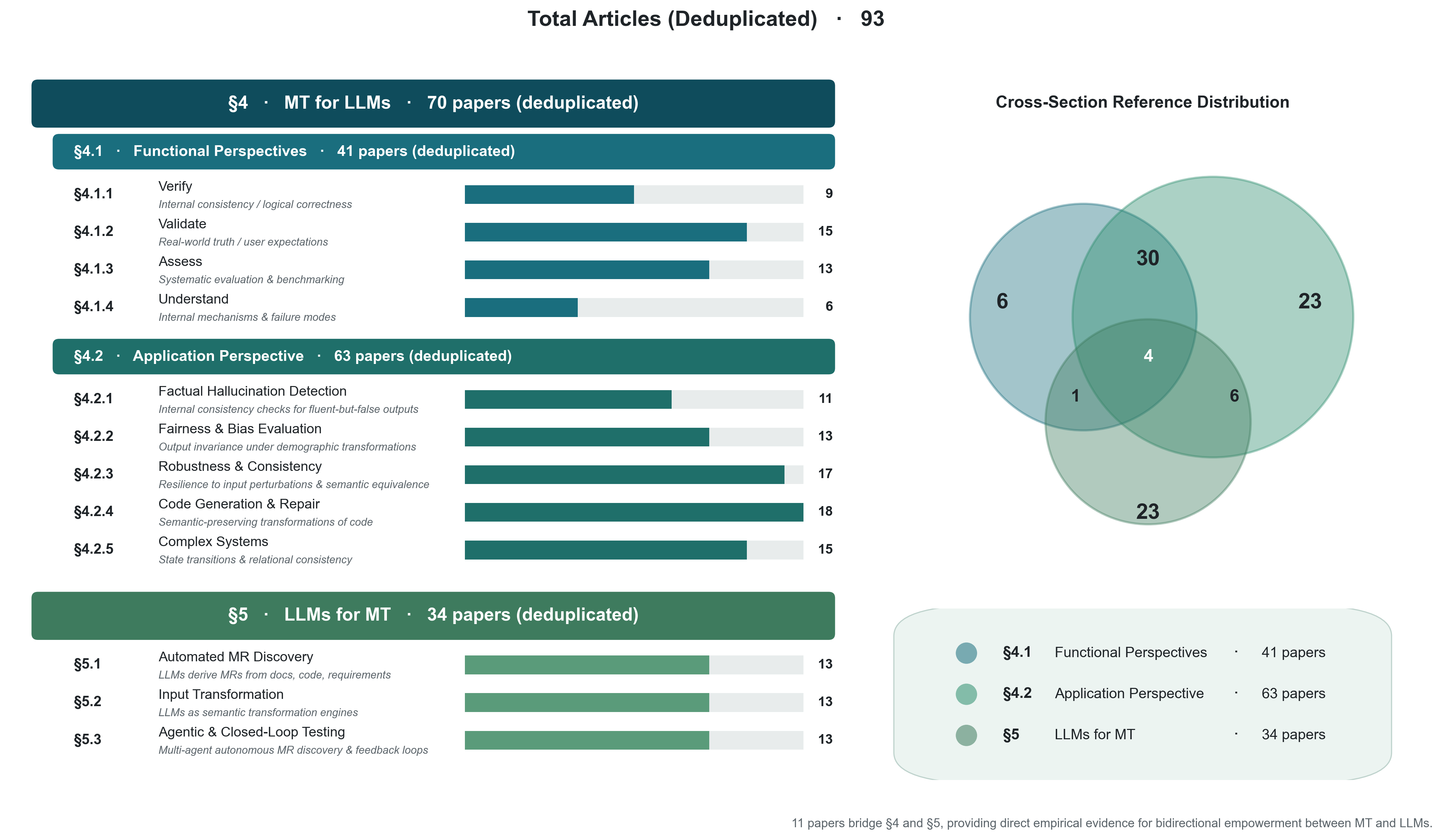}
    \caption{Taxonomy of the bidirectional relationship between MT and LLMs}
    \label{fig:Taxonomy}
\end{figure}

\subsection{Proposed Taxonomy}

Based on the 93 selected studies, we constructed a hierarchical taxonomy to systematize the bidirectional empowerment between MT and LLMs. As illustrated by Figure~\ref{fig:Taxonomy}, our taxonomy is structured around two complementary directions, MT for LLMs and LLMs for MT, which align with the core research questions RQ1 and RQ2. These categories are not intended to be mutually exclusive: a single study may instantiate multiple functional roles or address multiple quality attributes, while the taxonomy provides complementary lenses for organizing the literature.

\begin{enumerate}
    \item \textbf{MT for LLMs (Assuring Large Language Models):} This branch focuses on using MT as a rigorous, oracle-alleviating framework for LLMs. The literature is analyzed from two primary angles: the functional objectives of applying MT (what role it plays) and the concrete quality attributes it targets (what is being tested). From a functional perspective, MT serves to:
    \begin{itemize}
    \item \textbf{Verify} internal consistency and logical correctness, such as checking semantic invariance across semantically related input transformations.
    \item \textbf{Validate} alignment with real-world truths, user expectations, and domain constraints, including factual consistency, retrieved evidence, and other externally grounded requirements.
    \item \textbf{Assess} capabilities systematically for benchmarking, by evaluating model robustness, fairness, and other key attributes.
    \item \textbf{Understand} internal mechanisms and failure modes, exemplified by probing the behavioral fidelity of compressed models.      
    \end{itemize}
    From a quality attribute and application perspective, MT is instantiated to evaluate:
    \begin{itemize}
    \item \textbf{Factual Hallucination:} Detecting factually incorrect or unsupported generations by checking semantic, logical, temporal, or context-grounded consistency across related inputs and outputs.
    \item \textbf{Fairness \& Bias:} Uncovering social, demographic, intersectional, and domain-specific biases by testing whether outputs remain invariant or change appropriately under fairness-relevant perturbations.
    \item \textbf{Robustness \& Consistency:} Assessing stability under semantics-preserving transformations, adversarial or noisy perturbations, repeated executions, and production-like stress conditions.
    \item \textbf{Code Generation \& Repair:} Evaluating code-related LLM tasks, including code generation, completion, repair, summarization, and Text-to-SQL, by applying semantics-preserving transformations to prompts, programs, specifications, or generated artifacts.
    \item \textbf{Complex Systems - Agents \& Dialogue:} Testing stateful and interactive LLM-based systems through MRs over conversation histories, agent actions, tool use, environment states, and execution trajectories.
  \end{itemize}
  \vspace{0.5em}
    \item \textbf{LLMs for MT (Automating the Metamorphic Testing Lifecycle):} This branch leverages LLMs' generative and reasoning capabilities to automate some of the traditionally labor-intensive phases of MT.
    \begin{itemize}
     \item \textbf{Automated MR Discovery:} Using LLMs to extract, identify, and formalize candidate MRs from unstructured documentation, requirements, legal or domain rules, source code, existing test suites, and domain-specific patterns.
    \item \textbf{Input Transformation \& Test Implementation:} Using LLMs to synthesize executable transformation functions, generate semantically valid follow-up test cases, implement domain-specific test artifacts, and support semantic input/output patterns or translations.
     \item \textbf{Agentic \& Closed-Loop Testing:} Deploying LLM-based agents in collaborative, reflective, or feedback-driven workflows to generate, execute, evaluate, and refine metamorphic tests, using MR violations or consistency analyses to support self-correction and iterative improvement.
    \end{itemize}
\end{enumerate}

In addition to the hierarchical categories, Figure \ref{fig:Taxonomy} also summarizes the quantitative distribution
of the 93 deduplicated primary studies across the taxonomy. Overall, the current literature is more
heavily concentrated on the MT for LLMs direction, which includes 70 studies, whereas 34 studies
fall under the LLMs for MT direction. This imbalance suggests that, to date, MT has been more
widely explored as a quality assurance technique for LLMs than LLMs have been explored as an
automation mechanism for MT. Within MT for LLMs, the functional perspective
covers 43 papers,
while the application perspective covers 63 papers.
The larger number of papers in the application perspective indicates that existing work is often
driven by concrete LLM quality concerns and deployment scenarios. Within LLMs for MT, the 34
papers are distributed across three automation-oriented phases: automated MR discovery, input
transformation and test implementation, and agentic or closed-loop testing, showing that LLMs
are beginning to support multiple stages of the MT lifecycle rather than only isolated test-generation
tasks.

It should be noted that these counts are not mutually exclusive. Because a single study may address
multiple functional objectives, application scenarios, or automation phases, the sums of subcategory
counts may exceed the number of studies in the corresponding higher-level branch. The cross-section
distribution in Figure \ref{fig:Taxonomy} further highlights this overlap: some studies simultaneously contribute to
the functional analysis of MT for LLMs, the application-oriented evaluation of LLM systems, and
the automation of MT using LLMs. Therefore, the taxonomy should be interpreted as a multi-label
classification scheme rather than a partition of the literature. This quantitative overview provides
a bridge between the methodology and the subsequent survey sections: Section 4 reviews the
dominant body of work on MT for LLMs from both functional and application perspectives, while
Section 5 examines the emerging but rapidly growing line of work on LLMs for MT.

\section{MT for LLMs}


Building upon the conceptual roadmap in Figure~\ref{fig:empowerment}  and the taxonomy presented in Section 3.4, we now delve into a comprehensive survey of how MT is applied to LLMs. As shown in the left branch of Figure~\ref{fig:empowerment}, this direction is organized through two complementary lenses: a functional perspective that asks what role MT performs in the assurance lifecycle, and an application perspective that asks what specific attributes and application scenarios of LLM systems it targets.

\subsection{Functional  Perspectives}

Beyond serving as a pure oracle-alleviating testing technique, MT has evolved into a versatile analytical framework that supports multiple, distinct yet interconnected objectives in the quality assurance process for LLMs. The surveyed literature reveals that MT is systematically employed to achieve four core functional goals: verifying the internal logical and behavioral consistency of models, validating their alignment with external truths and constraints, systematically assessing their performance and robustness, and gaining a deeper understanding of their internal mechanisms and characteristic failure modes.

\subsubsection{Verify: Ensuring Internal Consistency and Logical Correctness}

The probabilistic nature of LLMs fundamentally challenges traditional software verification, which typically relies on strict adherence to formal specifications and expected outputs. In the absence of definitive test oracles, verifying an LLM shifts toward ensuring its internal consistency, reasoning stability, and logical correctness across varied contexts. MT addresses this by formalizing generally accepted behavioral boundaries—such as semantic invariance, symmetric stability, or logical entailment—across related executions. By doing so, it exposes contradictory or unstable outputs without requiring human-labeled ground truth.

A primary research stream leverages MT to verify the semantic invariance of LLMs under linguistic and contextual transformations. Pioneering this behavioral testing paradigm, CheckList~\cite{ribeiro2020beyond} introduces a comprehensive matrix of linguistic capabilities and test types—such as Invariance and Directional Expectation tests—to decouple testing from implementation, systematically exposing fundamental behavioral failures in commercial natural language processing (NLP) models. Building upon this foundation toward fully automated verification, LLMORPH~\cite{cho2025llmorph} operationalizes a robust suite of 36 MRs across multiple NLP tasks. It utilizes LLM-based transformations to generate follow-up inputs, automatically uncovering faulty behaviors and output inconsistencies at scale without the need for expensive labeled datasets. Moving beyond surface-level text tasks to complex scientific reasoning such as mathematics and physics, Curtò \&  Zarzà~\cite{decurto2025metamorphic} propose a metamorphic testing framework to evaluate the semantic invariance of reasoning traces. By applying structural, verbosity, and contextual transformations, their framework computes semantic similarity scores across intermediate reasoning steps, revealing a critical dissociation between an LLM's aggregate accuracy and its step-by-step reasoning fidelity. Further extending semantic invariance verification to educational Q\&A scenarios and content quality classification, Chan and Keung~\cite{chan2024symmetric} propose a symmetric MR for abbreviation–expansion transformation to verify the semantic invariance and internal consistency of LLMs, enhancing machine common sense and classification stability in educational Q\&A content quality tasks.

Beyond semantic equivalence, MT is rigorously applied to verify the logical soundness and temporal reasoning of generative models. To systematically expose fact-conflicting hallucinations, Drowzee-Temporal~\cite{li2025detecting} leverages metric temporal logic and logic programming to automatically synthesize complex, time-constrained question-answer pairs. It employs novel semantic-aware MRs to compare the logical structures of the LLM's reasoning steps against factual ground truth, effectively verifying whether the model's internal inference mechanisms violate established logical or temporal constraints.

Finally, the verification of internal consistency extends to code generation and specialized architectural components. In software engineering tasks, Chan et al.~\cite{chan2025effectiveness} evaluate the stability of code generation LLMs by applying symmetric MRs, such as prompt case transformation, duplication, and paraphrasing. Their approach verifies that semantics-preserving prompts consistently yield functionally equivalent source code, thereby exposing the models' brittle sensitivity to superficial prompt variations. In the Text-to-SQL domain, SQLHD~\cite{yang2025sqlhd} proposes a two-stage MT framework to verify LLM-generated schema mappings and SQL queries without requiring ground-truth SQL. It defines structure-aware and logic-aware MRs to cross-check source and follow-up outputs, detecting schema-linking hallucinations such as nonexistent tables or columns and logical-synthesis errors such as incorrect comparison ranges or extremum reasoning. Furthermore, MT has been adapted to verify the internal retrieval mechanisms of RAG pipelines. MeTMaP~\cite{wang2024metmap} verifies the correctness of vector matching in LLM-augmented systems by constructing sentence triplets based on word- and sentence-level MRs. It exposes widespread false vector matching problems, demonstrating that these underlying databases frequently erroneously prioritize structural similarity over genuine semantic correctness. Similarly, Tamanna et al.~\cite{tamanna2025chatgpt} propose CHIME, which applies MT to verify and repair RAG-based ChatGPT responses during software bug report understanding by generating semantically equivalent mutated queries and checking answer consistency. This work shows that MT can serve as an online verification mechanism for detecting inconsistent and potentially hallucinated LLM responses in domain-specific technical Q\&A systems.

\subsubsection{Validate: Confirming Alignment with Real-World Truths and User Expectations}

Validating the outputs of LLMs requires ensuring that they adhere to real-world truths, system constraints, and complex domain specifications. However, this validation process is inherently challenged by the absence of definitive ground-truth oracles and the black-box nature of LLMs. MT addresses this dilemma by asserting that a model's underlying knowledge and structural alignment should remain invariant or exhibit predictable variance under targeted, semantics-preserving transformations.

A primary research stream leverages MT to validate the factual alignment of LLMs and detect fact-conflicting hallucinations in open-domain natural language generation. Early zero-resource approaches, such as SelfCheckGPT~\cite{manakul2023selfcheckgpt}, operationalize this by stochastically sampling multiple responses and measuring informational consistency across these samples, in which significant output divergence indicates a hallucinated fact. Building on this principle, MetaQA~\cite{yang2025hallucination} employs explicit synonymy and antonymy MRs to transform base responses, identifying hallucinations by rigorously checking logical consistency without relying on external databases. To proactively induce diverse execution paths, DrHall~\cite{wu2025detectinga} operationalizes a comprehensive suite of questioning MRs and Answering MRs, utilizing multi-path voting to not only detect but also actively correct factual hallucinations. Moving beyond simple linguistic transformations, Drowzee~\cite{li2024drowzee} introduces a logic-programming-aided testing framework. It extracts factual triples from knowledge bases and applies fundamental logical reasoning rules—such as negation, symmetric, and transitive relations—to automatically generate MRs and related test cases, utilizing semantic-aware oracles to reliably capture discrepancies between the LLM's reasoning and real-world facts.

The validation of factuality has also been systematically extended to specialized architectures, such as RAG and complex conversational agents. In RAG systems, validating whether generated texts strictly align with retrieved evidence is critical. MetaRAG~\cite{sok2025metarag} formalizes a reference-free framework that decomposes LLM answers into atomic factoids, applies synonym and antonym substitutions to construct MRs, and validates these variants against the retrieved context to penalize unsupported claims at the span level. For conversational agents, Sensei~\cite{de2025automated} provides an end-to-end testing approach comprising a highly customizable user simulator and a domain-specific language to specify and check MRs across multi-turn dialogs. Similarly, to ensure the reliability of question-answering software, CQ2A~\cite{liu2025testing} employs context-driven question generation. By extracting entities and relationships to form ground truth answers, it utilizes LLMs to generate highly natural, context-covering questions, complemented by consistency checking to filter out false positives. To validate the intrinsic reliability of black-box models in classification tasks, researchers utilize MRs to assess prediction confidence and annotation quality. The perceived confidence score framework~\cite{salimian2025pcs} evaluates zero-shot classification confidence by applying semantics-preserving MRs, such as active-to-passive voice transformations and double negations. It aggregates label consistency across these variations to reliably estimate model confidence without requiring access to internal logits. In specific classification domains, such as Stack Overflow content moderation, Chan and Keung~\cite{chan2024validating} validate pretrained language models using semantics-preserving MRs—including text duplication and abbreviation replacement—to identify non-violation regions and support robust simulation testing for content quality.

Beyond natural language, validating LLM alignment with formal specifications presents unique challenges in software engineering tasks. 
Lin et al.~\cite{lin2026validating} propose MRSQLGen to validate whether LLM-generated SQL queries align with user intent by applying metamorphic prompting to the input question and comparing the execution results of the original and transformed queries. This work highlights that prompt-level transformations can serve as validation oracles for structured generation tasks where syntactically valid outputs may still violate user intent.
To address the lack of canonical solutions in automated code generation, Metamorphic prompt testing~\cite{wang2024validating} generates multiple program variants via prompt paraphrasing and cross-validates their functional alignment using automatically generated test inputs. In the context of code reading within big data environments, Li et al.~\cite{li2023evaluating} validate model comprehension by injecting structural MRs—such as altering numerical values or logical operators—to assess whether the LLM maintains a consistent understanding of the modified code constraints. Finally, to validate LLM responses in complex software troubleshooting workflows, ChatGPT incorrectness detector (CID)~\cite{tanzil2024chatgpt} targets technical software reviews. It employs an iterative "Enquirer-Challenger-Decider" architecture, transforming challenge questions based on MRs to systematically expose and detect inaccuracies in ChatGPT's technical reasoning. In a related software validation setting, Metamon~\cite{lee2025metamon} treats documentation as the specification and validates whether program behavior captured by regression tests conforms to that specification through LLM-mediated metamorphic prompts. This illustrates how MT can support alignment checking not only against external facts, but also against natural-language software specifications that are otherwise difficult to execute directly. In a related legal-critical setting, Gogani-Khiabani et al.~\cite{gogani2024technical} study the maintenance of tax preparation software under evolving IRS regulations and argue that tax-law updates can be translated into executable software artifacts with LLM support, while MRs derived from tax-policy properties remain essential for validating whether updated implementations preserve legal compliance.

\subsubsection{Assess: Systematic Evaluation and Benchmarking}

The systematic assessment of LLMs is fundamentally hindered by the limitations of static benchmarks, which frequently suffer from high annotation costs, limited diversity, and the pervasive risk of data contamination or data leakage. MT addresses these bottlenecks by providing a scalable, oracle-alleviating framework to dynamically assess and benchmark model capabilities across massive, unlabeled test groups.

A foundational research stream focuses on building unified, multi-dimensional benchmarking libraries for general NLP tasks. 
PromptOps~\cite{Sontesadisai2025PromptOps} advances this direction by providing a visual, data-flow-based tool for assessing LLM trustworthiness through MRs targeting robustness, fairness, and logical consistency. By supporting configurable perturbation operators and visual test reports, the tool lowers the barrier for practitioners to compare model behavior across prompts, perturbations, and LLM backends.
To thoroughly assess LLM qualities, METAL~\cite{hyun2024metal} formalizes MR templates covering essential quality attributes such as robustness, fairness, non-determinism, and efficiency, utilizing metrics like attack success rate combined with semantic text similarity to evaluate performance across varied generative tasks. Expanding on this, LLMORPH~\cite{cho2025llmorph} executes over half a million MGs across core NLP tasks such as natural language inference and sentiment analysis, demonstrating that MT can effectively complement traditional label-based testing by uncovering failures completely missed by static ground truths. To assess multimodal reasoning capabilities, MTEE~\cite{jiang2025metamorphic} proposes a unified MT approach targeting both textual and visual entailment tasks, using hypothesis-centric MRs driven by text decomposition to systematically detect and explain inference failures. 
Similarly, VRPTEST~\cite{li2023vrptest} evaluates large multimodal models under visual referring prompting by using MT to generate prompt variants that alter visual reference attributes such as color, font, shape, and position while preserving the intended task relation. This benchmark shows that visual prompting strategies can substantially affect LLM accuracy and that MT can support automated benchmark expansion without additional manual labeling.

Moving beyond standard NLP, systematic assessment is crucial in highly structured domains such as code generation and mathematical reasoning. To evaluate the true reasoning capabilities of instruction-tuned LLMs, Turbulence~\cite{honarvar2025turbulence} introduces the concept of question neighborhoods—parameterized sets of semantically equivalent programming prompts—and systematically evaluates LLMs by MT, which checks for consistency in code generation across these related input variations. In a similar vein, Sarker et al.~\cite{sarker2026assessing} formalize syntactic robustness for mathematical code generation, demonstrating that LLMs struggle to maintain consistent outputs when mathematical formulas in prompts undergo semantics-preserving syntactic transformations, and further propose a formula reduction technique to effectively mitigate these robustness failures.

As LLMs are integrated into complex, stateful architectures, assessment frameworks have evolved to target multi-turn dialogues, RAG, and autonomous agents. For conversational systems, Mortar~\cite{guo2025mortar} formalizes multi-turn MT by introducing dialogue-level transformations (e.g., round shuffling, reduction, or duplication) to assess context-dependency, operating entirely without the need for biased LLM-as-a-judge mechanisms. For autonomous agents, ReliabilityBench~\cite{gupta2026reliabilitybench} defines a three-dimensional reliability surface (consistency, robustness, and fault tolerance) and introduces action MRs—where correctness is based on end-state equivalence rather than textual similarity—to benchmark agent performance under production-like stress conditions. NoD-DGMT~\cite{wu2025detecting} further broadens agent assessment by focusing on decision optimality rather than task success alone. By applying diversity-guided MT in AI2-THOR simulator, it assesses whether embodied planners exhibit invariant optimality properties under position, action, condition, and scene transformations.

Finally, MT provides rigorous benchmarking tailored to specialized, high-stakes domains, such as manufacturing, healthcare, and education, where expected outputs are often ambiguous.
In the manufacturing sector, Li et al.~\cite{li2024empirical} construct a MT framework to systematically assess the accuracy and robustness of Chinese local LLMs across eight industrial scenarios, revealing significant ability gaps between local and global models when handling logical transformations versus industry-specific terminology. In the healthcare domain, Jaganathan et al.~\cite{jaganathan2025metamorphic} evaluate automated international classification of diseases (ICD) coding applications using parameterized MRs that simulate common clinical note errors such as typos, removed words, and acronym variations, highlighting the necessity of MT for ensuring the reliability of biomedical LLMs in safety-critical medical environments. Beyond industrial and healthcare scenarios, Haq and Cabot~\cite{ulhaq2026istqb} assess the software-testing knowledge of 60 multimodal LLMs using 30 international software testing qualifications board (ISTQB) certification exams and further apply semantics-preserving transformations to exam questions to distinguish robust conceptual understanding from memorization. 

\subsubsection{Understand: Uncovering Internal Mechanisms and Failure Modes}

Despite achieving remarkable performance on standard benchmarks, LLMs remain fundamentally opaque "black boxes". High accuracy scores frequently mask deeper structural deficiencies, such as data memorization, shallow reasoning, or severe representation gaps during model compression~\cite{xue2024exploring,awal2025metamorphic}. Traditional evaluation paradigms, which rely on static ground-truth comparisons, fail to explain why models succeed or how they inherently fail. MT addresses this opacity by acting as a systematic behavioral probe. By observing how model outputs shift—or remain invariant—across carefully crafted semantic and syntactic transformations, MT uncovers the internal reasoning mechanisms, representational fidelity, and hidden failure boundaries of LLMs without requiring explicit test oracles. MTEE~\cite{jiang2025metamorphic} provides a representative example of MT as an explanation mechanism rather than merely a failure detector. By decomposing a hypothesis into atomic sub-hypotheses and comparing which sub-hypotheses satisfy or violate the expected entailment relation, MTEE identifies critical entities, attributes, and relations responsible for the detected failures.

A primary research stream explores the internal representational fidelity of compressed LLMs. While knowledge distillation often produces smaller student models with accuracy comparable to their teachers,  MetaCompress~\cite{awal2025metamorphic} reveals that this is frequently a superficial pattern. By formalizing output-based MRs such as probability distribution similarity and high-confidence preservation, it performs a head-to-head comparison to expose deep discrepancies in behavioral fidelity, demonstrating that student models suffer up to 285\% greater performance drops under adversarial attacks. To mitigate this internal misalignment, MORPH~\cite{panichella2025metamorphic} incorporates MT into many-objective knowledge distillation by optimizing not only accuracy, model size, and efficiency, but also robustness to code transformations based on constructed MRs. Using prediction flips between original and semantically equivalent transformed code as the robustness signal, MORPH produces compressed models that are 47\% more robust than the state-of-the-art baseline.

Finally, MT is utilized to map the exact failure boundaries and data leakage of code generation models. To understand the impact of combinatorial transformations on failure detection, Hyun et al.~\cite{hyun2025search} employ multi-objective search algorithms on the METAL framework to optimize MR selection. Their empirical analysis uncovers silver bullet transformations (e.g., character-level graphical transformations and contextual synonym replacements) that expose fundamental cognitive vulnerabilities across diverse Text-to-Text tasks. In the domain of APR, MT-LAPR~\cite{xue2024exploring} applies MRs across token, statement, and block levels to reveal a critical internal mechanism: an LLM's robustness is strongly and positively correlated with code readability. Moreover, by evaluating LLMs on leakage-free datasets dynamically generated via MT, this framework successfully uncovers inflated performance caused by training data memorization, providing researchers with a deeper understanding of the models' true generalization boundaries. Extending this leakage-aware perspective, De Koning et al.~\cite{dekoning2026metamorphic} use MT as a behavioral probe to understand memorization in LLM-based APR. Their results show that performance degradation under semantics-preserving code transformations correlates with low negative log-likelihood, suggesting that MT can help distinguish genuine semantic repair capability from benchmark contamination.

\subsection{Application Perspective}

Complementing the functional perspective outlined above, the application of MT to LLMs can also be examined through the lens of the specific quality attributes and system types it targets. Driven by the established need for trustworthy software and the unique risk landscape of generative AI, researchers have instantiated MT to evaluate a diverse set of critical concerns. This section systematically categorizes the literature according to these target dimensions: factual hallucination, fairness and bias, robustness and consistency, code generation and repair, and the behaviors of complex agentic and dialogue systems. Each dimension introduces specific oracle challenges and demands tailored strategies for designing effective MRs. The following analysis highlights how MT operationalizes these abstract quality concerns into concrete, executable, relation-based testing practices, providing a structured view of its practical utility across the LLM application spectrum.

\subsubsection{Factual Hallucination Detection}

Hallucination remains one of the most critical reliability challenges for LLMs. While this phenomenon encompasses both faithfulness and factuality hallucinations, the latter---the generation of plausible-sounding but factually erroneous content---occurs more frequently in practice and poses a more severe threat to user trust~\cite{HuangYu2025, li2024drowzee, manakul2023selfcheckgpt}. Traditional evaluation methods, relying on static benchmarks or external knowledge bases, often suffer from information obsolescence and incompleteness~\cite{li2024drowzee,sok2025metarag}.  MT addresses these limitations by verifying the internal consistency of a model's knowledge and reasoning. It assumes that a hallucinating model will fail to maintain logical or semantic stability across a series of related inputs.

A primary research stream leverages semantic consistency to construct MRs, assuming that a model's grasp of true facts remains invariant against semantic transformations.  Early zero-resource approaches, such as SelfCheckGPT~\cite{manakul2023selfcheckgpt}, operationalize this by stochastically sampling multiple responses, in which significant output divergence indicates a hallucinated fact. However, passive sampling often fails to expose deeply embedded false beliefs. To actively provoke inconsistencies, DrHall~\cite{wu2025detectinga} devises six base and three composite MRs—including chain of thought, multilingual voting, and external knowledge augmentation—to systematically induce different execution paths in black-box LLMs, leveraging the instability of hallucinated answers under varied reasoning trajectories. Similarly, MetaQA~\cite{yang2025hallucination} implements a structured transformation generation and verification pipeline using synonymous and antonymous transformations. In the software engineering domain, CID~\cite{tanzil2024chatgpt} detects erroneous software reviews by rephrasing input queries and evaluating the semantic consistency of the model's subsequent answers. Despite these advancements, semantic consistency MRs share a fundamental limitation: they are prone to false negatives if the LLM suffers from consistent hallucinations---confidently and repeatedly generating the same incorrect factual assertions across all linguistic variations. Hallucination detection has also been extended to structured generation. MRSQLGen~\cite{lin2026validating} focuses on intent-violating hallucinations in Text-to-SQL, where generated SQL is executable but semantically misaligned with the user question, and uses metamorphic prompt variants plus execution-result relations such as equivalence, subset, and superset to expose such errors.

To address fact-conflicting hallucinations against established world knowledge, Drowzee~\cite{li2024drowzee} extracts factual triples from knowledge bases (e.g., Wikipedia) and utilizes logic programming to derive new facts via basic reasoning rules, such as negation, symmetric, inverse, and transitive relations. It then applies MT to construct test cases and evaluate responses.  Recently, this framework has been extended to temporal logic~\cite{li2025detecting}, incorporating metric temporal logic and Prolog to generate queries with quantitative timing constraints.

The widespread integration of RAG architectures has introduced unique reliability challenges, particularly extrinsic hallucinations where generated responses deviate from or conflict with the retrieved external context. To systematically address the generation phase of RAG pipelines, MetaRAG~\cite{sok2025metarag} introduces a reference-free, black-box MT framework. This approach initially decomposes the generated answers into indivisible atomic factoids. It subsequently applies controlled MT transformations to produce synonymous and antonymous variants for each factoid. MetaRAG then verifies these variants against the retrieved context to compute an overall hallucination score.

Finally, MT has been integrated into specialized architectural deployments. For complex technical analysis, CHIME~\cite{tamanna2025chatgpt} employs context-free grammar to parse technical components like stack traces, then leverages MT to iteratively generate follow-up questions to ensure the consistency of the model's technical reasoning. In educational contexts, DomainProbe~\cite{wei2025pilot} helps students identify hallucinations by prompting the LLM to extract and explain topical keywords; if these term-explanation pairs are identified as inconsistent, the output is flagged as untrustworthy, effectively mitigating the test oracle problem in self-learning AI applications.

\subsubsection{Fairness \& Bias Evaluation}
LLMs are being rapidly integrated into a myriad of software applications, which may introduce a number of biases, such as gender, age, or ethnicity~\cite{morales2023automating}. Because LLMs lack a clear test oracle and often produce stochastic outputs, MT provides a principled way to assess individual fairness without requiring exhaustive oracles, by checking output invariance under semantically neutral transformations.

A significant body of research focuses on frameworks for social bias detection. Early works explore the automated generation of test suites to assess potential biases in LLMs~\cite{morales2023automating}. To rigorously evaluate these systems, METAL~\cite{hyun2024metal} defines MR templates covering essential quality attributes, explicitly including fairness. It specifically evaluates whether the results produced by LLMs vary significantly based on the user's demographic characteristics. Meta-Fair~\cite{romero2025meta} comprises three components: MUSE for test case generation, GENIE for execution across multiple LLMs, and GUARD-ME, which employs LLMs as judges to identify inconsistencies in the results. To overcome the structural rigidity of template-based methods, GenFair~\cite{srinivasan2025genfair} leverages techniques such as equivalence partitioning, mutation operators, and boundary value analysis to generate source test cases. This approach improves fairness testing by producing linguistically diverse, realistic, and intersectional inputs.

To standardize these evaluations, CAFFE~\cite{parziale2025toward} formalizes LLM fairness test cases through explicitly defined components, including prompt intent, conversational context, input variants, expected fairness thresholds, and test environment configurations.
To detect violations beyond standard testable facets of ethics, Reeq~\cite{ma2025reeq} draws inspiration from reflective equilibrium, a modern reasoning method in moral and political philosophy. 
Using a sentence diversity-based approach, Giramata et al.~\cite{giramata2025efficient} demonstrate that prioritizing MRs improves fault detection rates compared to random prioritization, while reducing the time to the first failure.

Moving beyond single-dimension bias and general text evaluation, recent research addresses intersectional biases. Reddy et al.~\cite{reddy2025metamorphic} emphasize intersectional biases involving multiple overlapping attributes. They show that tone-based analysis captures subtler linguistic variations than traditional sentiment-based metrics, making it more sensitive to fairness violations. In natural language inference tasks, Li et al.~\cite{li2024detecting} design MRs targeting demographic indicators, such as sex, race, age, and socioeconomic status, to discern bias. Furthermore, the utility of MT is not only limited to detection. Salimian et al.~\cite{salimian2025bias} link testing with mitigation by using MRs to generate diverse, bias-inducing samples for model fine-tuning, which significantly enhances bias resiliency.

Fairness evaluation has also been extended to specific model architectures and vertical domains. In RAG systems, Oliveira et al.~\cite{oliveira2025fairness} show that minor demographic variations can break up to one-third of MRs, indicating that the retrieval component in a RAG pipeline can itself introduce or amplify bias.  In the medical domain, Jaganathan et al.~\cite{jaganathan2025metamorphic} evaluate the robustness and fairness of automated ICD coding programs that use BioMed LLMs, demonstrating that predictions can be adversely impacted by removing or modifying demographic details. Finally, Ruan et al.~\cite{ruan2024revealing} expose the fairness issues in Text-to-Image generation models by employing MT transformations, such as entity replacement and entity attribute enhancement, using names with regional characteristics.

\subsubsection{Robustness \& Consistency}

Robustness and consistency are fundamental requirements for evaluating LLMs. Robustness characterizes a model's resilience to input perturbations, while consistency ensures coherent responses across logically equivalent contexts or repeated invocations~\cite{decurto2025metamorphic,gupta2026reliabilitybench}. Semantic invariance evaluates whether a robust reasoning agent can maintain output stability when presented with semantically equivalent inputs subjected to various MRs~\cite{decurto2025metamorphic}.

Foundational frameworks have been established to systematically evaluate these attributes across general NLP tasks using MT. 
PromptOps~\cite{Sontesadisai2025PromptOps} instantiates robustness and consistency testing through nine perturbation types, including typo injection, synonym replacement, temporal context insertion, and coreference rewriting. Its preliminary evaluation on BoolQ shows that commercial LLMs such as GPT-4o and Gemini-2.0-Flash can still flip yes/no answers under subtle input perturbations, with observed answer changes ranging from small but non-negligible rates to over 30\% in some settings.
METAL~\cite{hyun2024metal} structures this evaluation by defining MRs as modularized metrics, employing templates such as equivalence (comparing output equivalence given original and perturbed inputs) and discrepancy (expecting different outputs for semantic-altering transformations). Similarly, LLMORPH~\cite{cho2025metamorphic, cho2025llmorph} serves as an automated testing tool that utilizes a comprehensive catalog of 36 MRs. It uncovers faulty behaviors through semantics-preserving transformations, operating without human-labeled data.  A complementary line of work evaluates robustness at the level of domain knowledge rather than task outputs. Haq and Cabot~\cite{ulhaq2026istqb} transform ISTQB certification questions using MRs such as synonym substitution, neutral sentence insertion, option reordering, and irrelevant-option injection, and compare model performance before and after transformation. This setting shows that metamorphic transformations can reveal whether LLMs genuinely understand software testing concepts or merely rely on memorized question formulations.

Beyond general NLP, robustness validation has been customized for domain-specific applications. For multimodal robustness, VRPTEST~\cite{li2023vrptest} studies whether LLMs behave consistently under controlled visual referring prompt transformations, including changes to reference color, shape, font, and position. The results reveal that seemingly superficial visual prompt choices can lead to large accuracy variations, highlighting the need for MT-based evaluation of multimodal prompt sensitivity. In healthcare, automated ICD coding applications using BioMed LLMs are evaluated for robustness and fairness using parameterized MRs that simulate common mistakes in clinical notes, such as typos or missing acronyms~\cite{jaganathan2025metamorphic}. Additionally, for Q\&A websites like Stack Overflow, semantics-preserving MRs are utilized to validate pretrained language models for content quality classification, identifying non-violation regions to support simulation MT~\cite{chan2024validating}. In a related educational Q\&A setting, Chan and Keung~\cite{chan2024symmetric} define a symmetric abbreviation--expansion MR that checks prediction consistency across semantics-preserving textual variants. While the transformed samples are used for data augmentation, the MR itself serves as a robustness-oriented validation criterion by requiring stable predictions under trivial linguistic changes. 

Robustness and consistency are also critical in LLM-based code generation. From the perspective of code intelligence, the systematic review by Asgari et al.~\cite{asgari2025metamorphic} confirms that MT has become a dominant paradigm for evaluating whether deep code models remain stable under semantics-preserving program transformations. The review also highlights that current robustness assessments are concentrated on Java, Python, and C/C++, motivating broader evaluation across languages and generation-oriented model architectures. The Turbulence benchmark~\cite{honarvar2025turbulence} systematically evaluates instruction-tuned LLMs by generating neighborhoods of related programming questions from natural language question templates.  This approach reveals reasoning gaps where models fail to generalize across an entire question neighborhood. To validate the stability of code generation outputs, symmetric MRs assert that code predictions should remain unaffected when the semantic meaning of a prompt is preserved~\cite{chan2025effectiveness}. Furthermore, in contexts involving formal specifications, syntactic robustness is evaluated by applying semantics-preserving syntactic transformations to mathematical formulas within prompts~\cite{sarker2026assessing}.

Finally, special attention is given to complex systems and autonomous agents. For RAG systems, MeTMaP~\cite{wang2024metmap} detects false vector matching problems by constructing sentence triplets, operating on the principle that semantically similar texts should match and dissimilar ones should not. To measure inconsistencies in an LLM's knowledge of the world, KonTest~\cite{akhond2025llm} leverages a knowledge graph to construct semantically equivalent queries, utilizing metamorphic and ontological oracles to reveal knowledge gaps.
For autonomous agents, ReliabilityBench~\cite{gupta2026reliabilitybench} evaluates agent reliability under production-like stress conditions across three dimensions: consistency under repeated execution, robustness to task perturbations, and fault tolerance under infrastructure failures.

\subsubsection{Code Generation \& Repair}

The application of LLMs to software engineering has revolutionized tasks such as code generation, completion, and APR. However, these models often suffer from the test oracle problem and security vulnerabilities.  MT has emerged as a vital technique for validating these black-box systems by defining semantics-preserving transformations on source code or natural language specifications to verify the consistency of the generated outputs.

A recent systematic literature review~\cite{asgari2025metamorphic} underscores the growing diversity of MT transformations in this field, reporting identifier renaming and dead code insertion as the most prevalent methods, while advocating for broader coverage across generative tasks. 
A primary area of application is APR and defect detection, where the goal is to fix buggy code without introducing new faults. MT-LAPR~\cite{xue2024exploring} introduces a comprehensive framework defining nine MRs across token, statement, and block levels. Addressing the risk of data leakage in APR benchmarks, CodeCocoon~\cite{demetamorphic} applies LLM-based synonym generation techniques to generate natural variants of defects. In a similar leakage-aware setting, De Koning et al.~\cite{dekoning2026metamorphic}
apply semantics-preserving metamorphic transformations to Defects4J and GitBug-Java, showing that performance
drops under transformation correlate with model familiarity signals and can reveal memorization-driven
APR success. However, deploying semantics-preserving transformations requires caution. A recent study on defect detection tools revealed a major issue: many publicly shared transformations inadvertently alter code semantics, thereby severely complicating their practical reuse~\cite{hort2025semantic}.

For code generation and completion tasks, MT can be used to ensure that models maintain structural and semantic consistency.
CCTEST~\cite{li2023cctest} targets code completion systems using program structure-consistent  transformations, repairing erroneous outputs by selecting the most average completion among all generated variants.
To validate code without canonical solutions, Metamorphic prompt testing~\cite{wang2024validating} creates program variants via prompt paraphrasing and cross-validates their execution outputs using random fuzzed inputs.
In the structured code generation setting, SQLHD~\cite{yang2025sqlhd} applies two-stage MT to detect hallucinations in LLM-based Text-to-SQL generation without requiring ground-truth SQL. It verifies schema-linking artifacts and generated SQL queries through structure-aware and logic-aware MRs, exposing invalid schema references and logically inconsistent clauses such as erroneous comparison ranges or extremum operations. 
Complementing SQLHD, MRSQLGen~\cite{lin2026validating} introduces metamorphic prompting for validating LLM-generated SQL queries. Instead of transforming SQL artifacts directly, it rewrites the natural-language prompt using hallucination-type-specific MRs and checks execution-result consistency across the original and metamorphic queries to detect intent-violating hallucinations without ground-truth SQL.
Expanding to prompt-level robustness, Turbulence~\cite{honarvar2025turbulence} evaluates question neighborhoods to systematically uncover anomalies in an LLM's ability to generalize. Similarly, other studies employ symmetric MRs  to expose output instability in code generation models, emphasizing the need for enhanced semantic comprehension~\cite{chan2025effectiveness}. Furthermore, for prompts containing mathematical formulas, LLMs frequently exhibit syntactic robustness failures. Thus, introducing a pre-processing step that mathematically reduces and simplifies formulas can significantly mitigate these vulnerabilities~\cite{sarker2026assessing}.
Moving beyond post-generation validation, some frameworks embed MRs directly into the synthesis pipeline.  For example, CodeMetaAgent~\cite{akhond2025llm} leverages MRs as proactive semantic operators to refine task specifications and synthesize robust test cases, transforming MT from a passive validation tool into an active driver of code generation. 

MT is also applied to assess code comprehension and summarization. To address outdated documentation, Metamon~\cite{lee2025metamon} generates MRs that compare an LLM's understanding of documented intent against actual code behavior captured via regression tests, thereby revealing semantic discrepancies.  Similarly, Khatib et al.~\cite{khatib2026examining} evaluate code summarization by injecting targeted behavioral changes into the code to check if the LLM updates its summary accordingly.
To improve reliability, it constructs metamorphic LLM queries by transforming assertions into their logical counterparts and aggregating repeated LLM judgments through self-consistency, achieving a precision of 0.72 and a recall of 0.48 on 9,482 documentation-test pairs.
Furthermore, in big data contexts, MRs based on prompt and relational operator modifications have been used to expose the foundational code reading boundaries of models like ChatGPT~\cite{li2023evaluating}.

Finally, MT evaluates the security and internal fidelity of code LLMs. CHTs~\cite{tan2025coverage} automates the detection of harmful content generation during code transformations. By measuring output damage, it exposes critical flaws in content moderation mechanisms.  Regarding model compression, MetaCompress~\cite{awal2025metamorphic} adapts MT to assess the behavioral fidelity of knowledge distillation. It demonstrates that while a smaller student model may achieve accuracy similar to its teacher, it frequently fails to replicate the teacher's internal representations under behavior-preserving MRs.

\subsubsection{Complex Systems: Agents \& Dialogue}

As LLMs evolve from static text generators into complex architectures comprising interactive dialogue and autonomous agents, the testing paradigm must shift. In such environments, correctness transcends simple input-output mappings; it depends instead on maintaining contextual state transitions, handling non-deterministic behaviors, and coordinating multi-agent workflows.  MT addresses the oracle problem within these dynamic systems by defining relations over conversation histories, agent actions, and system states.

\textbf{Multi-Turn Dialogue Systems.} Conversational systems exhibit high complexity due to their reliance on dialogue history and context. Masserini~\cite{masserini2026multi} outlines a multi-level testing plan for conversational AI, spanning from service-interaction to full multi-agent integrations, incorporating MT to automatically generate test cases based on semantics-preserving transformations. At the dialogue level, traditional isolated inputs are insufficient. Mortar~\cite{guo2025mortar} formalizes MT for stateful interactions by using dialogue-level transformations to verify correctness through context-preserving and context-altering MRs. To rigorously test these interactions, Sensei~\cite{de2025automated} provides a customizable user simulator driven by conversation profiles to generate meaningful interactions, enabling the assessment of functional correctness across dialogues. {CQ$^2$A}~\cite{liu2025testing} ensures context coverage via context-driven question generation, extracting entities and relationships to serve as ground truth. To mitigate hallucinations within technical dialogues, CHIME~\cite{tamanna2025chatgpt} employs query transformation and MT to verify responses. Furthermore, complex dialogue systems acting as recommenders or RAG pipelines are evaluated for robustness via rating shifting~\cite{khirbat2024metamorphic} and assessed for demographic bias using set equivalence MRs~\cite{oliveira2025fairness}. MTF~\cite{Henry2025MTF} provides an open-source MT framework for validating LLM-based web services, including dialogue-oriented applications such as the Mei-chan virtual companion agent. The framework separates input-pattern generation from test execution and supports reusable text-oriented MRs, enabling follow-up inputs such as paraphrases, synonym substitutions, and tone changes to be checked through semantic similarity-based output patterns.

\textbf{Autonomous Agents and Action Consistency.} For autonomous agents operating in non-deterministic workflows such as booking flights and manipulating web document object models (DOMs), textual output evaluation is inadequate. ReliabilityBench~\cite{gupta2026reliabilitybench} evaluates LLM agent reliability under production-like stress by introducing action MRs, where equivalence is strictly based on end-state equivalence rather than text similarity. Complementing end-state reliability, NoD-DGMT~\cite{wu2025detecting} targets the optimality of embodied-agent decisions by detecting cases where agents complete tasks successfully but inefficiently. It defines MRs over positions, actions, conditions, and scenes, and reports violations when follow-up scenarios yield lower-cost trajectories than the original plan, thereby extending MT from functional agent correctness to non-functional planning optimality.
Acknowledging the inherent non-determinism of agentic workflows, AgentAssay\cite{bhardwaj2026agentassay} shifts from binary verdicts to a stochastic test semantics with three-valued probabilistic outcomes (PASS, FAIL, INCONCLUSIVE), employing behavioral fingerprinting to map execution traces into compact vectors. Furthermore, for agents interacting with complex web environments, ASSURE~\cite{gao2025assure} presents a modular automated testing framework for AI-powered browser extensions, evaluating agent actions via semantic equivalence and security boundary MRs (e.g., hidden text manipulation).

\textbf{Domain-Specific Multi-Agent Systems.} As agents collaborate in critical domains, their emergent complexity requires systematic validation.  
TEMPLEs~\cite{Towey2025TEMPLEs} extends the application of MT-mediated LLM evaluation to higher education by framing teaching evaluation as an oracle-deficient task and using multiple personality-aware LLM evaluators as synthetic peer reviewers. In this framework, a master agent synthesizes diverse feedback while an MT verification agent checks whether expected pedagogical relations, such as constructive alignment between learning outcomes and assessments, are preserved.
AutoMT~\cite{liang2025automt} targets autonomous driving systems by employing a multi-agent framework where an M-Agent extracts MRs from traffic rules in Gherkin syntax, a T-Agent analyzes visual scenarios, and an F-Agent synthesizes valid follow-up test cases. In the legally critical domain, SYNEDRION~\cite{gogani2025llm} simulates software development teams for tax preparation software. A dedicated MetamorphicAgent collaborates with tax-expert and coder agents to synthesize software and generate higher-order MRs. These relations evaluate the rate of change across multiple taxpayer profiles (e.g., progressive tax brackets) to uncover systematic discrepancies.

\section{LLMs for MT: Automating the Testing Lifecycle}

While Section 4 demonstrated how MT serves as a quality assurance mechanism for LLMs, this section turns to the reverse direction in the bidirectional framework shown in Figure~\ref{fig:empowerment}, using LLMs to automate the MT lifecycle. The widespread adoption of MT in traditional software engineering has historically been hindered by an automation bottleneck. The emergence of LLMs offers a transformative solution to this bottleneck~\cite{molina2025test}. By leveraging their semantic understanding, code generation, and reasoning capabilities, LLMs are increasingly employed as intelligent assistants or agents to automate the MT lifecycle. This section surveys the methodologies where LLMs empower MT, categorized into MR discovery, test implementation, and agentic closed-loop testing.

\subsection{Automated MR Discovery}

The identification of valid MRs has traditionally been the most labor-intensive bottleneck in the application of MT, often requiring deep domain expertise to discern necessary properties from software specifications. LLMs have emerged as a powerful tool to overcome this barrier by automating the extraction of MRs from diverse software artifacts, ranging from unstructured natural language documentation to existing codebases~\cite{molina2025test}.

A primary line of research focuses on bridging the gap between informal requirements and formal specifications. Tsigkanos et al.~\cite{tsigkanos2023large} pioneered the use of LLMs to parse the user manuals of complex scientific software (e.g., storm-water management models).
Their extended study~\cite{Tsigkanos2023VariableDiscovery} formalizes this idea as an end-to-end LLM-based workflow for discovering I/O variables from scientific software manuals and demonstrates that LLMs can substantially automate the prerequisite variable-discovery step for subsequent MR construction. Extending this capability to full MR derivation, Shin et al.~\cite{shin2024towards} propose a framework that utilizes LLMs to interpret requirements documents and synthesize MRs in natural language. Crucially, their approach further instructs the model to translate these textual descriptions into a domain-specific language named SMRL. Similarly, in highly regulated domains like taxation, researchers have demonstrated that LLMs can translate complex legal texts into logic-based metamorphic specifications via few-shot in-context learning, significantly reducing the manual effort required to encode domain rules~\cite{srinivas2023potential}. 

Beyond documentation, existing software artifacts such as test suites and source code serve as rich resources for MR discovery. Xu et al.~\cite{xu2025automated} introduced MR-Scout, an approach that mines implicit MRs from developer-written test cases in open-source projects. By analyzing test assertions and method invocations, MR-Scout identifies test cases that essentially validate properties (e.g., asserting that formatting text as bold should not alter its content) and synthesizes them into parameterized, reusable MRs. MR-Coupler~\cite{xu2026mrcoupler}  advances source-code-based MR discovery by identifying functionally coupled method pairs through signature commonality, function calls, and state interactions, and then prompting LLMs to infer MRs and generate concrete metamorphic test cases from these pairs. 
In the mobile augmented reality (AR) domain, Bose et al.~\cite{bose2025llms} further study LLM-assisted MR identification by asking models to determine whether predefined AR-specific MRs apply to given source code snippets. It treats MR applicability detection as a subproblem of automated MR discovery and shows that multi-agent debate improves the stability and correctness of such judgments across MR rephrasings.

The viability of these LLM-driven discovery methods has been substantiated through empirical studies of increasing scale. Luu et al.~\cite{luu2023chatgpt} first reported an experience study on using ChatGPT to generate candidate MRs for nine target systems. Their results show that ChatGPT can propose correct MRs, including for systems that had not previously been tested with MT, but a large portion of the generated candidates are vague, unjustifiable, or incorrect, underscoring the continuing need for expert validation. Building on such early evidence, Zhang et al.~\cite{zhang2025can} evaluated the performance of models like GPT-3.5 and GPT-4 across 37 diverse software systems. Their findings confirm that LLMs can generate a substantial number of legal and correct MRs, and notably, can discover extra MRs that human testers had previously overlooked. However, the study also cautions that LLMs produce a non-negligible number of plausible-sounding but incorrect candidates. 

To mitigate this issue of unconstrained generation and reduce human refinement efforts, recent advancements advocate for injecting structured domain knowledge into the LLM's prompting process. 
SVPrompt-MR~\cite{Huang2025SVPromptMR} strengthens this direction by introducing a self-verification-based prompt construction strategy for MR identification. By combining role specification, MT concept introduction, structured MR output, domain-specific self-checking, and expert-feedback-based revision, the method covered all 44 expert-identified MRs and discovered 24 additional MRs in a thermal-hydraulic safety design program.
Building upon earlier experience reports in autonomous driving~\cite{zhang2023automated}, Zhang et al.~\cite{zhang2025enhancing} proposed a human-AI hybrid framework for the CARLA simulator that mitigates unconstrained generation by embedding predefined MR patterns into LLM prompts. This pattern-driven constraint, coupled with expert review, effectively narrows the search space, minimizes invalid MR candidates, and exposes system-level defects from natural language scenarios.

\subsection{Input Transformation \& Test Implementation}

Once an MR is conceptually defined, the testing process faces the practical hurdle of input transformation -- the construction of an executable mechanism capable of converting a source test case into a follow-up test case that satisfies the input relation. Traditionally, this phase sometimes represented a significant automation bottleneck, requiring testers to manually write complex, domain-specific  transformation scripts or rely on limited hard-coded input pairs. Recent advances have leveraged LLMs to overcome this barrier, fundamentally changing the implementation paradigm from manual scripting to automated synthesis.

A primary direction in this domain involves using LLMs to synthesize executable transformation code from static examples or abstract specifications. In many existing software repositories, developers encode MRs implicitly within test cases using hard-coded input pairs (e.g., manually calculating a date one day later) without implementing a reusable transformation function. Addressing this limitation, MR-Adopt~\cite{xu2024mr} treats the derivation of input relations as a programming by example task. By analyzing a single seed input-output pair, the framework employs an LLM to infer the intended logical relation---such as incrementing a date object---and synthesizes a generalized Java function to perform this transformation dynamically. This approach utilizes data-flow analysis to refine the generated code, successfully recovering executable transformations for over 72\% of incomplete MRs in open-source projects. Whereas MR-Adopt focuses on deducing reusable input transformation functions from implicit test pairs, MR-Coupler~\cite{xu2026mrcoupler} generates complete metamorphic test cases by combining functional-coupling analysis, LLM-based MR reasoning, input augmentation, and mutation-analysis-based validation. This illustrates a broader trend from generating isolated transformations toward synthesizing executable, validated MT artifacts end to end. Parallel to recovering implicit relations, Shin et al.~\cite{shin2024towards} demonstrated that LLMs can bridge the gap between natural language requirements and test execution. Their framework utilizes few-shot prompting to translate textual MR descriptions directly into executable MRs written in a specialized domain-specific language.

While synthesizing code works well for deterministic logic, domains involving natural language or complex semantics often require a different approach: employing the LLM itself as a semantic  transformation engine. For transformations where writing a deterministic script is infeasible---such as paraphrase this sentence while retaining sentiment or rewrite this code to use a different loop structure---the generative capabilities of LLMs offer a robust solution. LLMORPH~\cite{cho2025metamorphic, cho2025llmorph} formalizes this by distinguishing between simple function-based transformations (e.g., introducing typos) and complex LLM-based transformations. It utilizes models like Hermes-2 via few-shot prompting to implement sophisticated input  transformations, such as rewriting text with different styles or dialects. This method produces grammatically correct and semantically valid test cases that traditional rule-based transformation fails to generate. In static analyzer testing, StaAgent~\cite{nnorom2025staagent} uses LLMs as a semantic mutation engine to generate compilable Java mutants that preserve the behavior of seed bug programs while introducing structural changes such as dead stores, unreachable branches, and equivalent loop rewrites. These mutants instantiate follow-up test cases for MT and expose analyzer rules that are overly brittle to semantics-preserving program variations.

Complementing such LLM-driven transformations, MTF~\cite{Henry2025MTF} operationalizes reusable libraries of input and output patterns for LLM-based services, including multilingual word-level transformations, Japanese-specific paraphrases, and application-specific transformations. Its validation study further shows that the quality of input transformations and semantic output comparators can vary substantially across languages and edge cases, underscoring the need to validate MR libraries themselves.
Similarly, CodeMetaAgent~\cite{akhond2025llm} treats MRs as semantic operators. It prompts LLMs to systematically transform problem descriptions (e.g., via logical inversion or procedural decomposition), creating diverse reasoning paths that force the model under test to generalize beyond specific phrasing. Beyond textual and code inputs, DILLEMA~\cite{masudi2023dillema} extends LLM-assisted input transformation to vision-based MT by using an image captioning model and an LLM to identify modifiable attributes, generate alternatives, and produce task-preserving captions. These captions are then combined with the spatial structure of the original image through a diffusion model, enabling realistic follow-up image generation under MRs that preserve the original class or semantic labels. OBsmith~\cite{jiang2026obsmith} extends LLM-assisted test implementation to JavaScript obfuscator testing by using LLMs to generate program sketches that encode JavaScript-specific constructs, idioms, and corner cases; these sketches are then instantiated into executable programs and used to test whether obfuscation preserves program behavior.

Furthermore, LLMs facilitate input transformation in specialized and highly structured domains by handling complex syntactic translations. In the context of database testing, QTRAN~\cite{lin2025qtran} addresses the challenge of extending MT across different SQL dialects. It employs a two-phase LLM approach: first translating SQL queries from a source dialect (e.g., MySQL) to a target dialect (e.g., PostgreSQL) using RAG to map dialect-specific features, and then using a fine-tuned LLM to apply transformations to these translated queries. In the domain of embedded graphics, Hazott et al.~\cite{hazott2025llm} utilize LLMs to bridge the abstraction gap between high-level geometric properties (e.g., a line is equivalent to a rectangle of height 1) and low-level C++ library calls. The LLM synthesizes the specific firmware code required to render these shapes on virtual prototypes, enabling automated visual validation that was previously manual. Finally, frameworks like Turbulence~\cite{honarvar2025turbulence} automate the exploration of the input space through parameterization, using LLMs to instantiate question neighborhoods---sets of structurally similar coding problems with varying constraints---to systematically stress-test instruction-tuned models.

\subsection{Agentic \& Closed-Loop Testing}

The most advanced application of LLMs in the realm of MT involves the transition from tool-assisted and human-driven workflows to autonomous agents capable of managing the  testing lifecycle. Unlike the isolated tasks of MR discovery or input transformation discussed in previous sections, agentic frameworks establish a closed-loop system where LLMs not only generate test cases but also execute them, analyze the results, and autonomously refine the system under test or the test suite itself based on feedback from MT. 
The LLM-based workflow proposed by Cañizares et al.~\cite{canizares2026llmmtworkflows} operationalizes this closed-loop view through evaluator-fixer pipelines and conversational orchestration integrated into the Gotten MT framework. Across data-centre, autonomous-vehicle, and automata domains, the architecture generated semantically valid MRs and achieved over 99\% follow-up correctness with reasoning-oriented LLMs.
As an early example of feedback-driven closed-loop MT, Sudheerbabu et al.~\cite{sudheerbabu2024iterative} optimize the hyperparameters of metamorphic transformations through an online generative feedback loop. It illustrates the broader generative-AI-for-MT paradigm, where MR-check outcomes and system outputs are fed back to adaptively generate more fault-revealing follow-up tests for an industrial control system.

This paradigm shift is best exemplified by multi-agent orchestration, where complex testing objectives are decomposed into specialized roles handled by distinct LLM agents. In the domain of autonomous driving systems, AutoMT~\cite{liang2025automt} replaces manual scenario design with a collaborative architecture comprising three agents: an M-Agent that parses unstructured traffic laws into MRs, a T-Agent that analyzes the context of driving scenarios, and an F-Agent that synthesizes photorealistic follow-up test cases via RAG. This division allows the system to autonomously translate natural language specifications into executable metamorphic tests without human intervention. A similar collaborative pattern is observed in the legal domain with Synedrion~\cite{gogani2025llm}, a framework for developing tax preparation software. Synedrion employs a dedicated metamorphic agent that works alongside coder and tax-expert agents. By continuously generating higher-order metamorphic counterexamples derived from tax statutes, the metamorphic agent provides feedback that the coder agents use to iteratively refine software logic, ensuring compliance with complex regulatory constraints. 
Earlier work on LLM-assisted tax software maintenance~\cite{gogani2024technical} also sketches a closed-loop architecture in which LLM-generated candidate updates are ranked, checked by MRs, and then refined through feedback prompts when violations are detected. 
StaAgent~\cite{nnorom2025staagent} further demonstrates the use of LLM-based agents to automate MT for static analyzers. It decomposes the workflow into seed generation, code validation, mutation generation, and analyzer evaluation agents, where semantically equivalent mutants serve as follow-up inputs and inconsistent analyzer reports indicate rule implementation defects.

Beyond mere fault detection, these agentic frameworks increasingly utilize MRs as semantic operators for self-correction, effectively transforming MT from a testing technique into a runtime quality enhancement mechanism. CodeMetaAgent~\cite{akhond2025llm} operationalizes this by treating MRs as tools to transform problem descriptions and test inputs systematically. Rather than accepting a single output, the agent explores diverse reasoning paths triggered by these  transformations; it then analyzes the consistency of the resulting code solutions to identify errors and self-correct its output without external ground truth.  OBsmith~\cite{jiang2026obsmith} also demonstrates a feedback-driven testing loop in which failures and bug reports are analyzed by an LLM-based agent and then used to synthesize new sketches that generalize the observed obfuscator failure patterns. This turns discovered failures into reusable test-generation knowledge and illustrates how MT pipelines can become progressively more targeted.

Finally, the concept of closed-loop testing has extended into the ethical alignment of LLMs through dialogic reflection. 
In a related work-in-progress setting, TEMPLEs~\cite{Towey2025TEMPLEs} proposes an iterative multi-agent workflow in which LLM sub-agents provide teaching evaluations, a master agent consolidates them, and an MT verification agent returns refinement requests when predefined MRs are violated. This design highlights the potential of MT not only for testing software behavior, but also for mediating reliability and consistency in LLM-supported human assessment workflows.
Reeq~\cite{ma2025reeq} simulates the philosophical process of reflective equilibrium by employing an external LLM agent to critique the target model's suggestions. When a transformation (e.g., swapping demographic attributes) reveals an ethical inconsistency, the agent provides a critique that drives the target model to reflect and adjust its response. This iterative cycle continues until equilibrium is reached, dynamically ensuring the model's behavior aligns with ethical standards. SVPrompt-MR~\cite{Huang2025SVPromptMR} illustrates a closed-loop pattern for MR discovery in which LLM-generated relations are validated through program execution, expert feedback, and iterative prompt refinement. This feedback-driven process shows how self-verification and human-in-the-loop review can be combined to improve the correctness, coverage, and novelty of LLM-generated MRs in high-stakes scientific domains.
ReliabilityBench~\cite{gupta2026reliabilitybench} addresses this by introducing action MRs, where the agent's behavior is evaluated not by textual similarity, but by the equivalence of the end-state achieved. Similarly, frameworks like Sensei~\cite{de2025automated} employ user-simulator agents to drive end-to-end conversations, verifying that chatbots maintain consistent states across complex, multi-turn interactions.

\section{Challenges \& Future Directions}

As shown in Sections 4 and 5, existing studies have demonstrated the feasibility of using MT to evaluate LLMs and LLM-based systems, as well as the potential of using LLMs to automate the MT lifecycle. However, these two directions also introduce new methodological and practical challenges. Accordingly, this section synthesizes the key challenges associated with both directions and outlines future research directions toward a more rigorous, scalable, and trustworthy closed-loop ecosystem between MT and LLMs.

\subsection{Validity and Trustworthiness of Metamorphic Relations}

The effectiveness of MT fundamentally depends on the correctness and usefulness of the adopted MRs. However, in the context of LLMs, defining a valid MR is substantially more difficult than in traditional numerical or algorithmic software. Many LLM tasks are open-ended, underspecified, and highly context-sensitive. An apparently reasonable relation, such as requiring paraphrased inputs to produce equivalent outputs, may not always hold, because subtle lexical, pragmatic, cultural, or task-specific differences can legitimately alter the expected response. This creates a central tension: if an MR is too weak, it may miss important failures; if it is too strong, it may produce false alarms.

This problem becomes even more critical in the emerging line of work on LLM-generated MRs. While LLMs can greatly reduce the manual effort of MR discovery, they may also produce relations that are plausible-sounding but logically invalid, incomplete, or overly ambiguous. In this sense, the traditional automation bottleneck of MT may be partially replaced by a validity bottleneck. Future work should therefore focus on mechanisms for MR assurance, including:
\begin{itemize}
    \item \textbf{Human-in-the-loop validation}, where domain experts review and refine candidate MRs generated by LLMs;
    \item \textbf{Constraint-based MR checking}, where candidate relations are tested against specifications, existing test suites, formal invariants, or known domain constraints before adoption;
    \item \textbf{Multi-model cross-verification}, where several LLMs or heterogeneous analyzers independently generate, critique, or validate MRs;
    \item \textbf{Confidence estimation for MRs}, so that generated relations can be ranked by plausibility, specificity, operational feasibility, and expected fault-detection utility.
\end{itemize}

More broadly, the community needs a clearer theory of what constitutes a good MR for LLM-based systems. Future research may develop quality criteria such as validity, necessity, discriminative power, semantic stability, domain coverage, transformation diversity, operational feasibility, and cost-effectiveness, together with benchmark datasets for evaluating MR quality.

\subsection{Data Leakage, Benchmark Contamination, and Evaluation Inflation}

A recurring concern in LLM evaluation is data leakage, namely the possibility that benchmark instances, templates, or close variants have already appeared in pretraining, instruction-tuning, or alignment corpora. This issue undermines the validity of traditional static benchmarks and can also affect MT if source inputs or generated follow-up cases remain too close to memorized examples. 

This challenge is particularly serious for code generation, program repair, and widely circulated public datasets. Even when MT generates new test cases, the transformations may preserve enough surface structure that memorization effects persist. Similarly, if an LLM is used to generate test cases for another LLM from the same model family, hidden training overlap, shared pretraining data, or similar inductive biases may distort the evaluation. Thus, future work should explicitly design leakage-resistant metamorphic evaluation protocols, including:
\begin{itemize}
    \item \textbf{Leakage-resistant metamorphic benchmarks}, where MGs are dynamically generated from abstract schemas rather than fixed public instances;
    \item \textbf{Transformation diversity controls}, ensuring that follow-up cases differ sufficiently in syntax, structure, context, and problem formulation;
    \item \textbf{Benchmark provenance analysis}, tracing whether source tasks, templates, or code fragments are likely to overlap with training corpora;
    \item \textbf{Cross-family evaluation}, using different model families for test generation, execution, and judgment to reduce shared bias;
    \item \textbf{Continual benchmark refresh}, since both models and public corpora evolve rapidly.
\end{itemize}

In this sense, MT should not only be used as a testing technique, but also as a basis for constructing more contamination-resilient evaluation ecosystems.

\subsection{Non-Determinism, Reproducibility, and Test Stability}

LLM-based systems often exhibit non-deterministic or unstable behavior. Even under similar prompts and decoding parameters, outputs may vary due to sampling strategies, backend changes, model updates, retrieval states, tool environments, or hardware-level effects. This characteristic complicates the interpretation of MT results: an MR violation may indicate a genuine defect, but it may also reflect ordinary stochastic variation. Conversely, a relation may appear to hold in a small number of executions while failing under repeated trials.

Therefore, a key open problem is how to define stable and statistically meaningful MRs for probabilistic systems. Existing studies often rely on one-shot executions or a small number of samples, which may be insufficient for robust conclusions. Future work should move toward statistical MT for LLMs, including:
\begin{itemize}
    \item \textbf{Repeated execution} over the same metamorphic group;
    \item \textbf{Distribution-level comparisons} rather than single-output comparisons;
    \item \textbf{Statistical significance testing} for MR violations;
    \item \textbf{Robustness profiles} that summarize failure probabilities across temperatures, prompts, retrieval configurations, and system settings;
    \item \textbf{Reproducibility protocols} that record model version, API configuration, prompt template, decoding parameters, retrieval corpus, tool state, and execution environment.
\end{itemize}

Such efforts are particularly important for commercial foundation models, whose weights, retrieval backends, moderation policies, and serving stacks may change over time without full transparency. A mature MT methodology for LLMs will need to treat temporal drift, stochasticity, and infrastructure variability as first-class testing concerns.

\subsection{Scaling MT to Complex LLM Systems}

Most existing research on MT for LLMs focuses on validating single-turn prompts or isolated model outputs. However, real-world LLM applications are becoming increasingly complex, involving composite system architectures such as RAG pipelines, tool calling, workflow orchestration, memory modules, browser automation, and multi-agent collaboration. In such systems, failures often do not originate from a single model response but arise from the dynamic interactions among multiple components—retrieval, planning, execution, memory updating, tool invocation, and environmental feedback. This presents a fundamental challenge for MT: its unit of analysis must expand from static textual outputs to a systemic perspective encompassing processes, states, and trajectories.

Although recent studies have attempted to apply MT to dialogue systems and autonomous agents, the related work remains fragmented and insufficient. Current methods are largely confined to single-turn dialogues or superficial textual perturbations, lacking unified modeling of system states and coherent analysis of cross-turn trajectories, and are also inadequate for supporting compositional testing of multi-component systems. For instance, in a multi-agent system, a slight perturbation to one role’s instruction may propagate through collaborative dynamics, leading to non-local effects; in an RAG pipeline, failures may stem from any stage—retrieval mismatch, evidence fusion, citation planning, context truncation, or response generation; and for autonomous agents, the appropriate unit of correctness should be the final world state, execution cost, satisfaction of safety constraints, or trajectory consistency, rather than mere surface-level response similarity.

Therefore, to address the evolving forms of LLM systems, future MT research should explore the following directions:
\begin{itemize}
    \item \textbf{State-aware MRs} for workflows, trajectories, memory states, and environment transitions;
    \item \textbf{Compositional MT}, where system-level MRs are decomposed into relations for retrieval, reasoning, tool invocation, memory update, and action execution;
    \item \textbf{Interaction-level coverage metrics} for conversations, tool calls, agent plans, and environment states;
    \item \textbf{Online MT}, where metamorphic checks are embedded during runtime monitoring rather than only offline evaluation;
    \item \textbf{Testing for emergent collective behaviors}, especially in collaborative, competitive, or adversarial multi-agent settings.
\end{itemize}

This direction is likely to become increasingly important as LLM-based systems evolve from standalone models into interactive socio-technical ecosystems.

\subsection{Cost, Efficiency, and Industrial Practicality}

Although LLMs can automate many phases of MT, this automation is not free. Generating MRs, synthesizing follow-up cases, executing MGs, and performing verification may incur substantial computational and engineering costs, especially when frontier commercial models are involved. Some workflows require multiple LLM calls per test case, repeated executions for statistical stability, or human review for MR validation, making large-scale deployment expensive and slow. 

This cost issue creates a practical trade-off between test thoroughness and industrial feasibility. If MT is too expensive, difficult to integrate, or hard to interpret, practitioners may hesitate to adopt it despite its conceptual appeal. 
Future research should therefore focus on cost-aware and usability-oriented MT strategies, including:
\begin{itemize}
    \item \textbf{Identifying cost-effective classes of MRs}, especially those with high fault-detection return under constrained testing budgets.
    \item \textbf{Prioritizing MGs}, especially when computational resources are limited.
    \item \textbf{Selecting appropriate model backends}, especially when smaller local models can replace larger proprietary models in the MT pipeline.
    \item \textbf{Reducing testing overhead}, especially through caching, batching, sampling, or approximate semantic checking without sacrificing reliability.
    \item \textbf{Presenting MT results in actionable forms}, especially for developers, auditors, and domain experts.
\end{itemize}

There is also a need for engineering support, including reusable MT toolkits, benchmark repositories, integration with model evaluation platforms, and documentation practices for test artifacts. Bridging the gap between research prototypes and deployable industrial solutions remains an important future task.

\subsection{Toward Standardized Taxonomies, Datasets, and Metrics}

The surveyed literature is rich and fast-growing, but still fragmented. Although this survey proposes a taxonomy for organizing the bidirectional relationship between MT and LLMs, broader community-level standardization is still lacking. Different studies use inconsistent terminology, different MR definitions, different evaluation metrics, and different notions of failure. Even basic reporting dimensions, such as model version, prompt template, decoding setting, relation checker, execution count, retrieval corpus, and tool environment, are not consistently documented. This fragmentation makes replication difficult and hinders cumulative scientific progress.

To address this, the field would benefit from standardization efforts in several directions:
\begin{itemize}
    \item \textbf{Shared taxonomies} for MT objectives, MR types, transformation operators, relation checkers, and failure modes in LLM systems;
    \item \textbf{Public repositories of validated MRs} across domains, together with metadata about assumptions, applicable tasks, and known limitations;
    \item \textbf{Common reporting standards} for experimental setup and reproducibility;
    \item \textbf{Standard metrics} for MR quality, fault-detection effectiveness, false-positive rate, execution cost, and reproducibility;
    \item \textbf{Shared benchmark suites} spanning text, code, multimodal, RAG, tool-use, and agentic tasks.
\end{itemize}

Such infrastructure would allow more rigorous comparison across methods and accelerate the transition from isolated case studies to a coherent research area.

\section{Conclusion}

In this paper, we presented a systematic survey of the bidirectional relationship between MT and LLMs. As LLMs continue to reshape software engineering, natural language processing, and intelligent system development, ensuring their reliability, robustness, and trustworthiness has become an increasingly important research challenge. Within this context, MT provides a particularly suitable testing paradigm because it alleviates the oracle problem and enables behavioral verification by checking expected relations among multiple related executions. At the same time, the emergence of LLMs creates new opportunities to automate and strengthen the MT workflow itself. This reciprocal interaction defines a promising and rapidly growing research area.

From the perspective of MT for LLMs, the surveyed literature demonstrates that MT has become an effective approach for evaluating a wide variety of LLM quality attributes. Compared with conventional static benchmark evaluation, MT offers a more flexible and dynamic mechanism for uncovering hidden defects in systems whose outputs are open-ended, probabilistic, and difficult to verify against a single ground truth. From the perspective of LLMs for MT, recent studies indicate that LLMs can significantly enhance the automation, scalability, and accessibility of MT, particularly in MR discovery, input transformation, test implementation, and agentic testing workflows. These advances suggest that LLMs can serve not only as targets of testing, but also as intelligent assistants in the testing lifecycle. Nevertheless, our review also shows that the field remains in a formative stage and faces several important open challenges.

Looking forward, future research should focus on several directions. More rigorous methods are needed for validating and ranking MRs, especially those produced with LLM assistance. Hybrid oracle mechanisms that combine symbolic rules, retrieval evidence, statistical analysis, and model-based judgment should be explored to improve verification reliability. The community would also benefit from contamination-resistant benchmark design, repeated-execution and distribution-aware testing protocols, and compositional MT methods for complex agentic systems. In parallel, practical toolchains, public MR repositories, and shared evaluation standards will be essential for moving the field from promising prototypes to mature scientific and industrial practice.


\appendix

\bibliographystyle{plain} 

\bibliography{sample-base}  

\end{document}